% ----------------------------------------------------------------------------
%
% Plain TeX.    
%
% For correct cross-references, run twice (as with LaTeX).
%
% ----------------------------------------------------------------------------
%
\message{ Cross-referencing macros, B. Davies, version 20 Jan 1995 }

\chardef\catamp=\catcode`@
\catcode`@=11 

% look for xrf file from previous run

\newif\if@xrf\@xrffalse   % this becomes true once the xrf file is set up
\def\l@bel #1 #2 #3>>{\expandafter\gdef\csname @@#1#2\endcsname{#3}}
\immediate\newread\xrffile
\def\xrf@n#1#2{\expandafter\expandafter\expandafter
  \csname immediate\endcsname\csname #1\endcsname\xrffile#2}
\def\xrf@@n{\if@xrf\relax\else%
  \expandafter\xrf@n{openin}{ = \jobname.xrf}\relax%
  \ifeof\xrffile%
    \message{ no file \jobname.xrf - run again for correct forward references }%
  \else%
    \expandafter\xrf@n{closein}{}\relax%
    \setbox0=\hbox{\catcode`@=11 \input\jobname.xrf \catcode`@=12}%
  \fi\global\@xrftrue%
  \expandafter\expandafter\csname immediate\endcsname%
  \csname  newwrite\endcsname\xrffile%
  \expandafter\xrf@n{openout}{ = \jobname.xrf}\relax\fi}

% general macros which define \<string> and \ref<string> from \order{string}
% eg. \order{thm} defines \thm{<label>} and \refthm{<label>}

\newcount\t@g

% we want \<string>#1 to call \n@@<string>#1:#2:> but we don't know if the 
% arguments #1 and/or #2 are empty. 
% furthermore, if they are both non-empty they are separated by a colon
% so we add a colon and a further delimiter > and pre-process via \n@<string>
% \n@<string>#1:#2> calls \n@@<string>#1:#2:> in all cases

\def\order#1{%
  \expandafter\expandafter\csname newcount\endcsname
  \csname t@ghd#1\endcsname\csname t@ghd#1\endcsname=0
  \expandafter\def\csname #1\endcsname##1{\xrf@@n\csname n@#1\endcsname##1:>}

% define \n@@<string>#1:#2:> which assignes the number (t@ghd.t@g) 
% t@g is new unless the label is already in use - then a warning is issued
% t@ghd must change as soon as the trigger for \numberby changes
% eg, \numberby{\section} changes t@@ghd and we compare this with t@ghd

  \expandafter\def\csname n@@#1\endcsname##1:##2:>%
    {\edef\t@g{\csname t@g#1\endcsname}\edef\t@@ghd{\csname t@ghd#1\endcsname}%
     \ifnum\t@@ghd=\t@ghd\else\global\t@@ghd=\number\t@ghd\global\t@g=0\fi%
     \ifunc@lled{@#1}{##1}\global\advance\t@g by 1% check if this label in use
       {\def\n@xt{##1}\ifx\n@xt\empty%              blank label - don't record
       \else\writ@new{#1}{##1}{\pret@g\t@ghead\number\t@g}\expandafter%
       \xdef\csname @#1##1\endcsname{\pret@g\t@ghead\number\t@g}\fi}%
       {\pret@g\t@ghead\number\t@g}%                expand out the number
     \else\def\n@xt{##1}%
       \w@rnmess#1,\n@xt>\csname @#1##1\endcsname%  warning message then expand
     \fi##2}\ord@r{#1}}             

% same as \order{string} but with no prefixes - see previous comments

\def\sporder#1{%
  \expandafter\def\csname #1\endcsname##1{\xrf@@n\csname n@#1\endcsname##1:>}

  \expandafter\def\csname n@@#1\endcsname##1:##2:>%
    {\edef\t@g{\csname t@g#1\endcsname}%
     \ifunc@lled{@#1}{##1}\global\advance\t@g by 1%
       {\def\n@xt{##1}\ifx\n@xt\empty%
       \else\writ@new{#1}{##1}{\number\t@g}\expandafter%
       \xdef\csname @#1##1\endcsname{\number\t@g}\fi}{\number\t@g}%
     \else\def\n@xt{##1}\w@rnmess#1,\n@xt>\csname @#1##1\endcsname%
     \fi##2}\ord@r{#1}}

% macros common to \order and \sporder

\def\ord@r#1{%

% first define \n@<string>#1:#2> which sorts out which arguments are empty

  \expandafter\def\csname n@#1\endcsname##1:##2>%
    {\def\n@xt{##2}\ifx\n@xt\empty%
     \expandafter\csname n@@#1\endcsname##1:##2:>\else%
     \expandafter\csname n@@#1\endcsname##1:##2>\fi}

% define t@g<string> to keep count of the t@g for this entity

  \expandafter\expandafter\csname newcount\endcsname
  \csname t@g#1\endcsname\csname t@g#1\endcsname=0

% define \ref<string> as <string>c@te
% the arguments are delimited by either comma or dash and this is sorted
% out by \each@rg which applies the macor <string>c@te to the arguments
% one at a time putting the delimiters in between the output

  \expandafter\def\csname ref#1\endcsname##1{%
     \expandafter\each@rg\csname #1c@te\endcsname{##1}}

% <string>c@te has more work to do determining if there is a second argument
% which will be delimited by a colon
% all cases are reduced to calling <string>cit@#1:#2:,

  \expandafter\def\csname #1c@te\endcsname##1:##2,%
    {\def\n@xt{##2}\ifx\n@xt\empty%
     \csname #1cit@\endcsname##1:##2:,\else%
       \csname #1cit@\endcsname##1:##2,\fi}

% finally define <string>cit@#1:#2:, 
% this either expands out to the reference or gives an error message
% if the label has never been defined

  \expandafter\def\csname #1cit@\endcsname##1:##2:,%
    {\def\n@xt{##1}\ifunc@lled{@#1}{\n@xt}%
      {\expandafter\ifx\csname @@#1\n@xt\endcsname\relax%
       \und@fmess#1,\n@xt>>>\n@xt<<%
       \else\csname @@#1\n@xt\endcsname##2\fi}%
     \else\csname @#1\n@xt\endcsname##2%
     \fi}}

% for dealing with list of arguments separated by commas and
% dashes, and applying a specified control sequence to each
% so we can have \ref{a,b}, etc 

\def\each@rg#1#2{{\let\thecsname=#1\expandafter\first@rg#2,\end,}}
\def\first@rg#1,{\callr@nge{#1}\apply@rg}
\def\apply@rg#1,{\ifx\end#1\let\n@xt=\relax%
\else,\callr@nge{#1}\let\n@xt=\apply@rg\fi\n@xt}

\def\callr@nge#1{\calldor@nge#1-\end-}
\def\callr@ngeat#1\end-{#1}
\def\calldor@nge#1-#2-{\ifx\end#2\thecsname#1:,%
  \else\thecsname#1:,\hbox{\rm--}\thecsname#2:,\callr@ngeat\fi}

% for writing labels to jobname.xrf

\def\writ@new#1#2#3{\xrf@@n\immediate\write\xrffile
  {\noexpand\l@bel #1 #2 {#3}>>}}

% for checking if labels are undefined

\def\ifunc@lled#1#2{\expandafter\ifx\csname #1#2\endcsname\relax}
\def\und@fmess#1#2,#3>{\ifx#1@%
  \message{ ** error - eqn label >>#3<< undefined - run again ** }\else
  \message{ ** error - #1#2 label >>#3<< undefined - run again ** }\fi}
\def\w@rnmess#1#2,#3>{\ifx#1@%
  \message{ Warning - duplicate eqn label >>#3<< }\else
  \message{ Warning - duplicate #1#2 label >>#3<< }\fi}

% to precede equation numbers by a taghead

\def\t@ghead{}
\newcount\t@ghd\t@ghd=0
\def\taghe@d#1{\gdef\t@ghead{#1}\global\advance\t@ghd by 1}

% define equation labels with an @qn

\order{@qn}

% now define \eqno(<label>), \leqno(<label>), \ref(<label>)
% also redefine \eqalignno & \leqalignno so that &() is picked up 

\let\eqno@@=\eqno
\def\eqno(#1){\xrf@@n\eqno@@\hbox{{\rm(}$\@qn{#1}${\rm)}}}

\let\leqno@@=\leqno
\def\leqno(#1){\xrf@@n\leqno@@\hbox{{\rm(}$\@qn{#1}${\rm)}}}

\def\ref#1{\xrf@@n{{\rm(}$\ref@qn{#1}${\rm)}}}

% the following only differs from plain TeX by picking up equation labels
% \@centering is the same name as in LaTeX for \centering

\newskip\@centering
\@centering=0pt plus 1000pt minus 1000pt
\def\eqalignno#1{\xrf@@n\displ@y \tabskip=\@centering
  \halign to\displaywidth{\hfil$\displaystyle{##}$\tabskip=0pt
   &$\displaystyle{{}##}$\hfil\tabskip=\@centering
   &\llap{$\eqaln@##$}\tabskip=0pt\crcr
   #1\crcr}}
\def\leqalignno#1{\xrf@@n\displ@y \tabskip=\@centering
  \halign to\displaywidth{\hfil$\displaystyle{##}$\tabskip=0pt
   &$\displaystyle{{}##}$\hfil\tabskip=\@centering
    &\kern-\displaywidth\rlap{$\eqaln@##$}\tabskip\displaywidth\crcr
   #1\crcr}}
\def\eqaln@#1#2{\relax\ifcat#1(\expandafter\eqno@\else\fi#1#2}
\def\eqno@(#1){\xrf@@n\hbox{{\rm(}$\@qn{#1}${\rm)}}}

	% the following hack allows plain TeX \eqalign to work in LaTeX

\ifx\@latexerr\undefined\else%        
  \def\eqalign#1{\null\,\vcenter{\openup\jot\m@th
    \ialign{\strut\hfil$\displaystyle{##}$&$\displaystyle{{}##}$\hfil
    \crcr#1\crcr}}\,}
\fi

% macro to make order use a prefix
% this is very general, for example the section and/or subsection numbers 
% may be used, but any command which is not outer will do via the call
% \numberby{\cs} or \numberby{\csa,\csb} 
% internally, \cs or \csa is called \s@ction and \csb is called \subs@ction
% \cs, \csa or \csb are redefined so that some preprocessing is done 
% since \numberby can be called more than once, we save their names 
% in \s@@ve and \subs@@ve, and restore them on a subsequent call to \numberby

% \numberby can have 0, 1 or 2 arguments, separated by commas
% 0 arguments means to stop using a t@g which we do by making it empty

\def\numberby#1{\xrf@@n
  \ifx\s@ction\undefined\else
  \expandafter\let\csname\s@@ve\endcsname=\s@ction\fi
  \ifx\subs@ction\undefined\else
  \expandafter\let\csname\subs@@ve\endcsname=\subs@ction\fi
  \numb@rby#1,>#1>}

% if 0 arguments, #1 #2 and #3 are all empty - blank t@ghd
% if 1 argument, it goes to #1 and #3 but #2 is empty - use n@by
% if 2 arguments, #2 has an extra comma but not #3 - use n@@by
 
\def\numb@rby#1,#2>#3>{\def\n@xt{#1}\ifx\n@xt\empty\taghe@d{}\else
  \def\n@xt{#2}\ifx\n@xt\empty\n@by#3>\else\n@@by#3>\fi\fi}

% the case, eg, of a section number as t@ghd
% the input command \cs is saved in \s@@ve and then redefined

\def\n@by#1>{\ifx\s@cno\undefined\expandafter\expandafter
  \csname newcount\endcsname\csname s@cno\endcsname
  \csname s@cno\endcsname=0\else\s@cno=0\fi
  \xdef\s@@ve{\expandafter\n@@me\string#1>}
  \let\s@ction=#1\def#1{\global\advance\s@cno by 1
  \taghe@d{\number\s@cno.}\s@ction}} 

\def\n@@me#1#2>{#2}

% the case, eg, of section.subsection numbers as t@ghd
% we must deal with two commands \csa and \csb

\def\n@@by#1,#2>{\ifx\s@cno\undefined\expandafter\expandafter
  \csname newcount\endcsname\csname s@cno\endcsname
  \csname s@cno\endcsname=0\else\s@cno=0\fi
  \ifx\subs@cst\undefined\expandafter\expandafter
  \csname newcount\endcsname\csname subs@cst\endcsname
  \csname subs@cst\endcsname=0\else\subs@cst=0\fi
  \ifx\subs@cno\undefined\expandafter\expandafter
  \csname newcount\endcsname\csname subs@cno\endcsname
  \csname subs@cno\endcsname=0\else\subs@cno=0\fi
  \xdef\s@@ve{\expandafter\n@@me\string#1>}
  \let\s@ction=#1\def#1{\global\advance\s@cno by 1
  \global\subs@cno=\subs@cst
  \taghe@d{\number\s@cno.}\s@ction} 
  \xdef\subs@@ve{\expandafter\n@@me\string#2>}
  \let\subs@ction=#2\def#2{\global\advance\subs@cno by 1 
  \taghe@d{\number\s@cno.\number\subs@cno.}\subs@ction}} 

% to reset counters

\def\numberfrom#1{\ifx\s@cno\undefined\else\n@mberfrom#1,>\fi} 
\def\n@mberfrom#1,#2>{\def\n@xt{#2}%
  \ifx\n@xt\empty\n@@f#1>\else\n@@@f#1,#2>\fi} 
\def\n@@f#1>{\s@cno=#1\advance\s@cno by -1} 
\def\n@@@f#1,#2,>{\s@cno=#1\advance\s@cno by -1%
  \subs@cst=#2\advance\subs@cst by -1} 

% macro to add additional prefix to numbers

\def\pret@g{}
\def\prefixby#1{\gdef\pret@g{#1}}

\catcode`@=\catamp

\message{ Numerically ordered citation macros, B. Davies, version 20 Jan 1995 }

\chardef\catamp=\catcode`@
\catcode`@=11 

% miscellaneous definitions

\ifx\bibrm\undefined\def\bibrm{\relax}\fi%     allow for different size
\def\l@ft{[}%                                  left bracket for number
\def\r@gth{]}%                                 right bracket for number
\def\g@p{\hskip0.2em}%                         space after commas and dashes
\def\d@sh{\hbox to 0.2em{}--\hskip0.2em}%      dash between entries in list
\newif\ifciteup\citeupfalse
\def\citeup{\citeuptrue%
  \def\g@p{\hskip0.1em}%                       spaces in upper list
  \def\d@sh{-}%                                dashes in upper list
  \def\l@ft{}%                                 dashes in upper list
  \def\r@gth{.}}%                              dashes in upper list

\newcount\r@fcount\r@fcount=0
\newcount\r@fcurr
\newcount\r@fmin
\newcount\r@fmax
\newcount\r@fone
\newcount\r@ftwo
\newtoks\c@toks
\newif\ifc@te\c@tefalse
\newif\ifr@fnext

% definition of \cite
% this must do a number of things
% i)   make a note of which labels are used in this \cite command and give them
%      numbers if it is their first appearance
% ii)  for each of citation, produce a skeleton \bibitem 
%      "Reference <lable> to be supplied"
% iii) produce a list of numbers generated by this \cite, 
%      sorted in ascending order and with proper punctuation
% iv)  format the output either as [..] with the optional argument last,
%      or as superscripts with the optional arguments omitted

\def\cite#1{\c@te#1>>}

\def\c@te#1#2>>{\ifx#1[\def\n@xt{}\else\def\n@xt{]{#1#2}}\fi\expandafter\c@@te\n@xt}

% note -- if the optional argument is present, #1 will be [ and #2 will be empty
% because \cite has undelimited parameter #1
% in this case \c@@te still needs to pick up the rest of the argument list -- 
% the next tokens still waiting to be digested (the [ has been eaten)
% otherwise \c@te already has everything in #1#2, and we just pass it on via \n@xt

% after this \c@@te knows the optional argument even if it is empty

% \s@p contains the punctuation required before the first entry of a group
% a group is of the form "n," or "n,n+1" or "n--m"
% there is nothing before the very first group, subsequently a comma
% initialise \s@p to {}

\def\c@@te#1]#2{\def\s@p{}\r@fmin=\r@fcount\r@fmax=0%
  \refn@te#2>>\r@fcurr=\r@fmin\advance\r@fcurr by-1\c@toks={\refc@te}%
  \ifciteup%
    \unskip$^{\the\c@toks}$\def\n@xt{#1}%
    \ifx\n@xt\empty\else\c@toks={#1}%
    \message{ Optional argument [\the\c@toks] in cite ignored }\fi%
  \else%
    \def\n@xt{#1}%
    \ifx\n@xt\empty[\the\c@toks]\else[\the\c@toks,\g@p#1]\fi%
  \fi}

% make a note of which references are cited and allocate number if required
% input list is separated by commas so we add one more to treat recursively

% after the recursive use of \refn@@te below, \r@fmin and \r@fmax will
% contain the interval in which cited numbers lie

\def\refn@te#1>>{\refn@@te#1,>>}

% strip any leading or trailing blanks by adding blank and delimiter >>
% \r@fnote is used as \r@fnote#1 >> but defined as \r@fnote#1#2 #3>>
% #1 is undelimited -- leading blanks ignored 
% #2 is everything else up to first blank
% this strips blanks fore and aft leaving label as #1#2
% embedded blanks are not allowed and cause a log message

% the reason for this elaboration is that it may be convenient to have the
% argument list of \cite spread over more than one line of input without
% using trailing % characters to suppress blanks

\def\refn@@te#1,#2>>{\r@fnote#1 >>%
  \def\n@xt{#2}\ifx\n@xt\empty\else\refn@@te#2>>\fi}% \refn@@te until list exhausted

% now for each cited number, define \r@fn<number> to by {C} rather than {}
% for later use by \refc@te to detect which numbers have been cited
% also adjust the interval [\r@fmin,\r@fmax] as each new number turns up

\def\r@fnote#1#2 #3>>%
  {\def\n@xt{#3}\ifx\n@xt\empty\else
    \message{ Embedded blanks not allowed in cite #1#2 #3}\fi%
  \ifunc@lled{r@f}{#1#2}%
    \global\advance\r@fcount by 1\r@fmax=\r@fcount\r@fcurr=\r@fcount%
    \expandafter\xdef\csname r@f#1#2\endcsname{\number\r@fcount}%
    \expandafter\gdef\csname r@ftext\number\r@fcount\endcsname%
    {{#1#2} Reference #1#2 to be supplied}%
  \else%
    \expandafter\r@fcurr=\csname r@f#1#2\endcsname\relax%
    \ifnum\r@fmin<\r@fcurr\else\r@fmin=\r@fcurr\advance\r@fmin by-1\fi%
    \ifnum\r@fmax<\r@fcurr\r@fmax=\r@fcurr\fi%
  \fi%
  \expandafter\expandafter\def\csname r@fn\number\r@fcurr\endcsname{C}}

\def\ifunc@lled#1#2{\expandafter\ifx\csname #1#2\endcsname\relax}

% loop through the interval [\r@fmin,\r@fmax] and cite as appropriate

\def\refc@te{\r@fnextfalse\def\s@ve{}%
  {\loop\ifnum\r@fcurr<\r@fmax\advance\r@fcurr by 1\c@tefalse%
  \expandafter\refc@@te\number\r@fcurr>>%
  \ifc@te\expandafter\refn@xt\number\r@fcurr>>\fi\repeat\s@ve}}

% \refc@@te examines \r@fn<number>
% if it is nonempty reset to {} and set \c@tetrue

\def\refc@@te#1>>{\ifnum#1=0%
    \edef\n@xt{}%
  \else
    \edef\n@xt{\csname r@fn#1\endcsname}%
  \fi%
  \expandafter\xdef\csname r@fn#1\endcsname{}%
  \ifx\n@xt\empty\else\relax\c@tetrue\fi}

% process next cited reference in the loop
% this is a bit tricky -- we want to use commas as separators except when
% there are groups of three or more consecutive numbers which should just
% appear as the first and last, separated by a dash
% we need to know (i) if this is the first group (ii) where we are in the group
% \r@fnextfalse flags the very first call to \refn@xt -- takes care of (i)
% \r@fone contains the first number of the current group and \r@ftwo keeps track 
% of where we are up to -- takes care of (ii)

\def\refn@xt#1>>{\ifr@fnext\ifnum\r@fone=\r@ftwo\r@second#1>>%
  \else\r@third#1>>\fi\else\r@fnexttrue\r@first#1>>\fi}

\def\r@first#1>>{\r@fnexttrue\r@fone=#1\r@ftwo=#1%
  \s@p\expandafter\relax\number\r@fone}%

% after the first number, the possible completion of this group is kept in 
% \s@ve until a non-consecutive number turns up, when \s@ve is passed on
% then \r@first is invoked again to start a new group

\def\r@second#1>>{\advance\r@ftwo by 1\def\n@xt{#1}%
  \expandafter\ifnum\n@xt=\number\r@ftwo%
  \edef\s@ve{,\g@p\expandafter\relax\number\r@ftwo}%
  \else\def\s@p{,\g@p}\r@first#1>>\fi}%

\def\r@third#1>>{\advance\r@ftwo by 1\def\n@xt{#1}%
  \expandafter\ifnum\n@xt=\number\r@ftwo%
  \edef\s@ve{\d@sh\expandafter\relax\number\r@ftwo}\def\s@p{,\g@p}\else%
  \def\s@p{,\g@p}\s@ve\def\s@ve{}\r@first#1>>\fi}%

% the rest is easy!!

% \bibit@m will produce formatted output in numerical order
% the space left for [..] is determined from the with of [<\the\r@fcount>]~
% there is a small further hanging indentation [0.5em] in our layout
% in case LaTeX is in use, the LaTeX definition is used for \bibit@m

\ifx\@latexerr\undefined%
  \newbox\n@mbox%                                   to get widest [..]
  \newdimen\n@mskip%                                fixed indentation for [..]
  \newdimen\h@ng\h@ng=0.5em%                        hanging indent amount
  \newdimen\@ldskip%		                              save the old values
  \newdimen\@ldindent%
  \def\bibit@m#1#2>>{{\leftskip=\n@mskip\advance\leftskip by \h@ng%
    \parskip=0pt\parindent=-\h@ng\frenchspacing%
    \hskip-\n@mskip\hbox to \n@mskip{\bibrm\l@ft\number\r@fcurr\r@gth\hss}%
    \ignorespaces#2\par\vskip\@ldskip%
    \parskip=\@ldskip\parindent=\@ldindent\leftskip=0pt}}
\else%
  \let\bibit@m=\bibitem%               LaTeX -- save its definition of \bibitem
\fi

% \bibitem will read the details of references and match with cited ones
% we can redefine it in LaTeX since we saved the old definition in \bibit@m

\def\bibitem#1#2\par{\bibitem@#1#2>>}%     ungroup #1 to get at first character
\def\bibitem@#1#2>>{\ifx#1[\bibitem@@#1#2>>\else\bibitem@@[]#1#2>>\fi}
\def\bibitem@@[#1]#2 #3>>{%
  \def\n@xt{#1}\ifx\n@xt\empty\else%
    \c@toks={#1}%
    \message{ Optional argument [\the\c@toks] in bibitem ignored }\fi%
  \bibitem@@@{#2}{#3}}
\def\bibitem@@@#1#2{%
  \ifunc@lled{r@f}{#1}\else%
    \expandafter\gdef\csname r@ftext\csname r@f#1\endcsname\endcsname{{#1} #2}%
  \fi}

% for producing the \jobname.bbl file

\newif\ifbbl@\bbl@false

\def\makebbl{\bbl@true%
  \expandafter\immediate\csname newwrite\endcsname\bbl@ % 
  \immediate\openout\bbl@ = \jobname.bbl%
  \def\writ@bbl##1>>{\c@toks={##1}%
    \immediate\write\bbl@{\noexpand\bibitem \the\c@toks \noexpand\par}}}

% produce the references in numerical order

\ifx\@latexerr\undefined%        
  \def\listreferences{%
    \setbox\n@mbox=\hbox{\bibrm\l@ft\number\r@fcount\r@gth~}\n@mskip=\wd\n@mbox%   
    \@ldskip=\parskip\@ldindent=\parindent%                save old values
    {\global\r@fcurr=0%
    \loop\ifnum\r@fcurr<\r@fcount\global\advance\r@fcurr by 1\relax%   
    \expandafter\expandafter\expandafter\bibit@m%
    \csname r@ftext\number\r@fcurr\endcsname>>\repeat}%
    \ifbbl@
      {\global\r@fcurr=0%
      \loop\ifnum\r@fcurr<\r@fcount\global\advance\r@fcurr by 1\relax%   
      \expandafter\expandafter\expandafter\writ@bbl%
      \csname r@ftext\number\r@fcurr\endcsname>>\repeat}\fi}
\else%
  \let\@nd=\end%
  \def\end#1{\def\n@xt{thebibliography}\def\n@@xt{#1}%
  \ifx\n@xt\n@@xt\listr@ferences{}\fi\@nd{#1}}%
  \def\listr@ferences{%
    {\global\r@fcurr=0%
    \loop\ifnum\r@fcurr<\r@fcount\global\advance\r@fcurr by 1\relax%   
    \expandafter\expandafter\expandafter\bibit@m%
    \csname r@ftext\number\r@fcurr\endcsname\repeat}%
    \ifbbl@
      {\global\r@fcurr=0%
      \loop\ifnum\r@fcurr<\r@fcount\global\advance\r@fcurr by 1\relax%   
      \expandafter\expandafter\expandafter\writ@bbl%
      \csname r@ftext\number\r@fcurr\endcsname>>\repeat}\fi}
\fi%

\catcode`@=\catamp

\message{Reference formatting macros, B. Davies, version 5 Oct 1996.}

% Standard macros for setting out
% automatic punctuation is provided so that groups may be empty

% \jnlitem has four arguments typeset as follows:
% #1 Roman -- authors, title of paper
% #2 Slanted -- name of journal
% #3 Bold -- volume number
% #4 Roman -- year, page numbers

% \bkitem has three arguments
% #1 Roman -- authors
% #2 Slanted -- name of book
% #3 Roman -- publisher, year, page numbers

\ifx\bibrm\undefined\def\bibrm{\relax}\fi
\ifx\bibit\undefined\def\bibit{\sl}\fi
\ifx\bibbf\undefined\def\bibbf{\bf}\fi

\chardef\catamp=\catcode`@
\catcode`@=11 

\newif\ifs@p
\def\jnlitem#1#2#3#4%
  {\frenchspacing{\bibrm#1}%
   \def\l@st{#1}\ifx\l@st\empty\s@pfalse\else\s@ptrue\fi%
   \def\l@st{#2}\ifx\l@st\empty\else%
   \ifs@p, \fi{\bibit#2}\s@ptrue\fi%
   \def\l@st{#3}\ifx\l@st\empty\else\ifs@p, \fi{\bibbf#3}\s@ptrue\fi%
   \def\l@st{#4}\ifx\l@st\empty\else\ifs@p, \fi{\bibrm#4}\s@ptrue\fi%
   \ifs@p.\fi\par}
\def\bkitem#1#2#3%
  {\frenchspacing{\bibrm#1}%
   \def\l@st{#1}\ifx\l@st\empty\s@pfalse\else\s@ptrue\fi%
   \def\l@st{#2}\ifx\l@st\empty\else%
   \ifs@p, \fi{\bibit#2}\s@ptrue\fi%
   \def\l@st{#3}\ifx\l@st\empty\else\ifs@p, \fi{\bibrm#3}\s@ptrue\fi%
   \ifs@p.\fi\par}

\catcode`@=\catamp

%
% abbreviations for commonly quoted journals
%

\def\AP{Adv. Phys.}

\def\CMP{Comm. Math. Phys.}

\def\JMP{J. Math. Phys.}
\def\JPA{J. Phys. A: Math. Gen.}

\def\JSP{J. Stat. Phys.}

\def\MPL{Modern Physics Letters}
\def\NC{Nuovo Cim.}

\def\NPB{Nucl. Phys. B}

\def\PA{Physica A}
\def\PD{Physica D}

\def\PLA{Phys. Lett. A.}

\def\PR{Phys. Rev.}

\def\PTP{Prog. Theor. Phys.}
\def\PTRSL{Phil. Trans. Roy. Soc. Lond.}

\def\ZPB{Z. Phys. B}

\input epsf.tex   
%
% ----------------------------------------------------------------------------
% Plain TeX, define page size and headings
% ----------------------------------------------------------------------------
%
\magnification 1200
\vsize=23truecm
\voffset=-.6truecm
\hsize=15.6truecm
\hoffset=.6truecm
\parskip=2pt
\nopagenumbers
\pageno=0
\headline={\ifnum\pageno=0{}\else \hdline\fi}
\def\hdline{\ifodd\pageno\rightheadline \else\leftheadline\fi}
\def\rightheadline{\tenrm\hfil{\it \rhead}\hfil\folio}
\def\leftheadline{\tenrm\hfil{\it \lhead}\hfil\folio}
%
% ----------------------------------------------------------------------------
% section and subsection headings LaTeX compatible
% ----------------------------------------------------------------------------
%
\catcode`@=11
\newif\ifnews@ct\news@ctfalse 
\parskip=2pt plus 2pt
\font\twelvebf=cmbx10 scaled 1200  
\font\twelveit=cmti10 scaled 1200
\newcount\secno\secno=0
\def\section#1#2\par#3\par
  {\vskip2cm\penalty-250\vskip-2cm\bigskip
  \if#1*\noindent{\bf#2}\else
  \advance\secno by 1\subsecno=0\news@cttrue
  \noindent\hbox to \parindent{\bf\number\secno.\hfil}{\bf#1}\fi
  \par\vskip-\parskip\medskip#3\news@ctfalse\par}
\newcount\subsecno\subsecno=0
\def\subsection#1#2
  {\ifnews@ct\else\smallskip\fi
  \if#1*\noindent{\sl #2.}\hskip1ex\else\advance\subsecno by 1 
  \noindent\number\secno.\number\subsecno\hskip1ex{\sl #1.}\hskip1ex\fi}
%
% ----------------------------------------------------------------------------
% abbreviations for this paper
% ----------------------------------------------------------------------------
%
\def\CTM{{\rm C\kern-.3pt T\kern-.3pt M}}   
\def\XYZ{{\rm X\kern-.3pt Y\kern-.3pt Z}}
\def\ABF{{\rm A\kern-.3pt B\kern-.3pt F}}
\def\AN#1{A^{(NS)}(#1)} 
\def\AR#1{A^{(R)}(#1)}
\def\BN#1{B^{(NS)}(#1)}
\def\BR#1{B^{(R)}(#1)}
\def\HN{H^{(NS)}}
\def\HR{H^{(R)}} 
\def\NS{Neveu-Schwarz}
\def\R{Ramond} 
\def\sn{\mathop{\rm sn}} 
\def\cn{\mathop{\rm cn}}
\def\dn{\mathop{\rm dn}}
\def\ns{\mathop{\rm ns}}
\def\nd{\mathop{\rm nd}}
\def\ds{\mathop{\rm ds}}
\def\tr{\mathop{\rm trace}}
\def\ip(#1){(#1)}
\def\0{0${}^\circ$}
\def\point(#1){\par\noindent\hbox to 16pt{%
  (#1)\hss\ignorespaces}\hangindent=16pt\hang}
\def\np{\par\noindent\ignorespaces}  
%
% ----------------------------------------------------------------------------
%
\catcode`@=12
%
% ----------------------------------------------------------------------------
%

\def\ltitle{A unified treatment of Ising model magnetizations}
\def\stitle{Ising model magnetizations}

\def\lauthor{Brian Davies${}^\dagger$ \cr
                 and\cr
             Ingo Peschel \cr}

\def\sauthor{Brian Davies and Ingo Peschel}

\def\laddress{Fachbereich Physik, Freie Universit\"at Berlin, \cr
  Arnimallee 14, D--14195 Berlin, Germany. \cr}

\def\paddress{${}^\dagger$~Permanent address: \cr
Department of Mathematics, School of Mathematical Sciences, \cr
Australian National University, Canberra, ACT 0200, Australia. \cr}

\def\labstract{
We show how the spontaneous bulk, surface and corner magnetizations
in the square lattice Ising model can all be obtained within one approach. 
The method is based on functional equations which follow from the properties 
of corner transfer matrices and vertex operators and which can be derived
graphically.
In all cases, exact analytical expressions for general anisotropy are
obtained.
Known results, including several for which only numerical computation was
previously possible, are verified and new results related to general
anisotropy and corner angles are obtained.}

%
% ----------------------------------------------------------------------------
% Typeset first page and eject
% ----------------------------------------------------------------------------
%
\baselineskip=18pt
\gdef\lhead{\sauthor}
\gdef\rhead{\stitle} 
\vglue 1cm
\centerline{\twelvebf\vbox{\halign{\hfil # \quad\hfil\cr\ltitle\crcr}}} 
\vglue 1cm
\centerline{\twelveit\vbox{\halign{\hfil # \quad\hfil\cr\lauthor\crcr}}}
\vglue 2cm
\baselineskip=12pt
\centerline{\vbox{\halign{\hfil # \quad\hfil\cr\laddress\crcr}}}
\vglue 1cm
{\narrower
\font\smr=cmr10 scaled 937
\font\smb=cmbx10 scaled 937
\noindent{\smb Abstract.}\hskip1ex\smr\labstract\par} 
\vfill
\centerline{To appear in Annalen der Physik}
\vfill
\noindent {\bf Keywords:} Ising model, magnetization, functional equations,
surfaces, corners.
\bigskip
\noindent Revised version, November 25, 1996
\vglue 1cm 
\noindent\vbox{\halign{# \hfil\cr\paddress\crcr}} 
\eject
%
% ----------------------------------------------------------------------------
% now for the actual body of the paper
% ----------------------------------------------------------------------------
%
\order{fig}%           automatic numbering of figures with \fig and \reffig
 
\numberby{\section}
%
%
%               Section 1
%
%
\section{Introduction}

The calculation of order parameters in solvable statistical models has by now a
rather long history.
It began with Onsager's result for the bulk magnetization of the square lattice
Ising model \cite{Ons49,Yang52}.
Quite some time later, McCoy and Wu obtained the order at a $(10)$ surface of
the same system \cite{MW67,MW73}, by which the general study of surface critical
behaviour began.
Again it took some time until it was realised that this behaviour depends on
the shape of the surface and that an interesting phenomenon, namely continuously
varying exponents, appears at corners \cite{Car83,BPP84}.
Thus Ising models with corners have been studied over the last 12 years
\cite{IPT93} but so far there has been only one completely analytical
calculation, namely for a square lattice with a 9\0\ corner formed by two
$(10)$ surfaces \cite{AT94}.

The calculations in all these cases can be (and have been) done using row transfer matrices.
One then has to find, for example, the matrix element of a spin operator between
two asymptotically degenerate eigenvectors of this matrix.
However, while this is rather simple for the $(10)$ surface \cite{Abr71,KP89}, 
especially in the anisotropic (Hamiltonian) limit \cite{Pes84},
it involves the solution of an integral equation for the bulk and the corner
problem \cite{KP89}.
Thus an alternative and possibly simpler method is desirable.

Such a technique has become available in the form of Baxter's corner transfer
matrix (\CTM) \cite{BaxBk}.
In fact, the bulk order parameters in most of the more complicated solvable
models have been obtained through it.
In this method, a square lattice is divided into quadrants and the \CTM\ is the
partition function of one such quadrant with the variables at the outer
boundary held fixed.
One then can calculate directly the expectation value of the central spin in
the form of a trace.
A basic feature is the simple structure of the \CTM\ eigenvalue spectrum in the
thermodynamic limit \cite{BaxBk} which leads to simple infinite product
formulae for the order parameters \cite{Bax85}.

While in the early \CTM\ calculations the spectrum was used directly a somewhat
different approach has been developed recently by the Kyoto group
\cite{DFJMN93,JMN93,Dav94}.
In this approach, so-called vertex operators play a fundamental r\^ole.
These are half-infinite lines of vertices in vertex models, or lines of
couplings in an Ising model, and hence can be viewed as half-infinite row
transfer matrices.
If they differ from the lattice couplings only in their anisotropy (spectral
parameter), then they have simple commutation relations with each other and with
the \CTM s.
This can be used to ``rotate'' such lines in a lattice and to obtain in this
way functional equations (also known as $q$-difference equations) for order
parameters or even correlation functions. 
(Actually the idea of moving lines about in a general way was introduced by
Baxter for the eight vertex model \cite{Bax78}, a concept called
``$Z$-invariance''.)
Such equations can be solved in some cases in a proper elliptic functions
parametrization and this leads to the same type of infinite product formulae as
are observed in cases where direct calculation is possible.

So far, this technique has been used mainly for obtaining some bulk order
parameters, e.g. in \ABF\ models \cite{ABF84,DJKMO87}.
Only recently, vertex models with straight surfaces have been considered,
and the spontaneous magnetization at a free end of the \XYZ\ spin chain has
been obtained \cite{JKKMW95}.
An additional ingredient needed here are boundary Yang-Baxter equations and the
corresponding $K$-matrices.

In this paper we want to consider the square lattice Ising model and
show how, using this approach, the various types of order parameters
can all be obtained in a unified way.
We recover all previously known results and derive several new ones. 
Although the Ising model is a special case of the eight-vertex or the \ABF\
models, we think that it is useful and warranted to treat it completely on its
own.

Thus we will first outline the technique of forming the functional equations
in a spin formulation and show how to recover once more the Onsager formula.
Then we describe how the procedure has to be modified for a system with a free
$(11)$ surface, i.e., one cut diagonally with respect to the couplings.
It turns out that, instead of complicated Yang-Baxter boundary equations, one
needs only a simple elementary relation here.
The structure of the functional equations remains essentially the same as in
the bulk and we obtain from it the magnetization in the first and second row.
We then apply the procedure to a 9\0\ corner formed by two $(11)$
surfaces and obtain the magnetization at three different corner points.
The solution for this corner is especially interesting in the critical region,
because the corner exponent $\beta_c$ depends on the anisotropy.
We obtain this dependence, which previously was inferred by rescaling onto
an isotropic system \cite{Car83,BPP84,IPT93}, directly from our analytical
formulae.

Finally we show that the method also works for a lattice with a $(11)$ surface
and a layered structure of the couplings perpendicular to it.
An appropriate choice of the layering then produces $(10)$ surfaces, so
that these can also be treated.
In particular, one can obtain the other type of 9\0 corner in this way
and derive the corresponding magnetization formula.
This completes our unified treatment.
A concluding section contains some additional remarks and an outlook.

%
%
%               Section 2
%
%
\section{Basic constructions}

\subsection{Parametrisation}
We consider an Ising model defined on a regular square lattice taken
on the diagonal, and parametrise the couplings using multiplicative
rapidity variables $\zeta_1$, $\zeta_2$ attached to lines running in the
horizontal and vertical directions. 
Except in section 6, we do not account for the horizontal and vertical rapidities
individually, simply writing $\zeta=\zeta_1/\zeta_2$.

\vbox{\bigskip
\centerline{\epsfbox{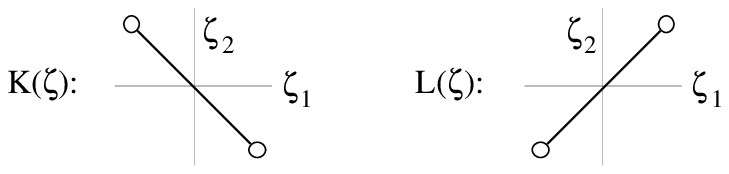}}
\bigskip
\centerline{Figure \fig{couplings}: Ising couplings $K$ and $L$.}
\bigskip}

There are two types of coupling as shown in figure \reffig{couplings}.
Our parametrisation uses the standard formulae \cite{BaxBk}
$$
\eqalign{\sinh2K&=-i\sn(iu),\cr
\cosh2K&=\quad\cn(iu),\cr}
\qquad
\eqalign{\sinh2L&=-i\sn(iI'-iu)\cr
\cosh2L&=\quad\cn(iI'-iu)\cr}
\hskip-4pt\eqalign{&=i\ns(iu)/k,\cr&=i\ds(iu)/k.\cr}
\eqno(KLdef)
$$
where $\sn$, $\cn$ are the usual Jacobian elliptic functions, of modulus
$$
k=1/(\sinh2K\sinh2L).
$$
This parameter, already introduced by Onsager, measures the temperature.
The quarter periods of the elliptic functions are $I$ and $I'$, where
$I=I(k)$ is the elliptic integral of the first kind and $I'=I(k')$ with
$k'=\sqrt{1-k^2}$.
The anisotropy of the couplings is measured by the variable $u$, or
equivalently by 
$$
\zeta=\exp(-\pi u/2I).
$$
Furthermore we will use the quantities
$$
q=\exp(-\pi I'/I),\qquad
x=q^{1/2},
$$
In these variables the crossing symmetry (rotation through 9\0) $L(u)=K(I'-u)$ 
becomes $L(\zeta)=K(x/\zeta)$.
In our calculations we shall particularly need an expression for 
$\tanh L(\zeta)$ in terms of infinite products.
It is
$$
\tanh L(\zeta)
={\ip(q^{1/2}/\zeta;q^2)\ip(q^{3/2}\zeta;q^2)
 \ip(-q^{3/2}/\zeta;q^2)\ip(-q^{1/2}\zeta;q^2)
 \over\ip(-q^{1/2}/\zeta;q^2)\ip(-q^{3/2}\zeta;q^2)
 \ip(q^{3/2}/\zeta;q^2)\ip(q^{1/2}\zeta;q^2)},
\eqno(tanhH)
$$
where $\ip(z;p)$ is the infinite product
$$
\ip(z;p)=\prod_{n=0}^\infty(1-zp^n).
$$
The identification \ref{tanhH} follows from standard elliptic function
identities, together with the fact that the Jacobi theta
function $H(u)$ has the infinite product expansion
$H(iu)=iq^{1/4}z^{-1/2}\ip(q^2;q^2)\ip(z;q^2)\ip(q^2/z;q^2)$,
where $z=\exp(-\pi u/I)$.
We note for later use the relation
$$
{H(iu)\over H(iu+I)}
=k'^{1/2}{\sn(iu)\over\cn(iu)}
=i{\ip(z;q^2)\ip(q^2/z;q^2)\over
 \ip(-z;q^2)\ip(-q^2/z;q^2)},
\qquad
z=\exp(-\pi u/I).
\eqno(theta)
$$
\subsection{Star-triangle relation}
Integrability depends on special properties of the Boltzmann weights,
particularly the star-triangle relation shown in figure \reffig{str1}.
The filled circle represents a spin which is summed out, while the scalar
factor is determined by the normalisation of the Boltzmann weights.
The condition on the rapidities in the three couplings is given by Baxter
\cite{BaxBk}; taking into account the present notation it is

\vbox{\bigskip
\centerline{\epsfbox{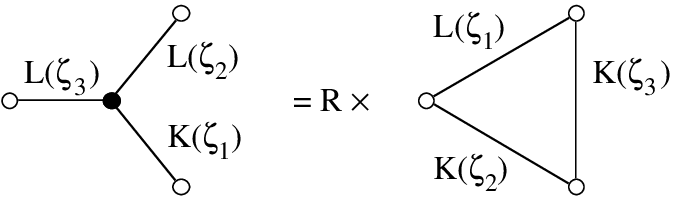}}
\bigskip
\centerline{Figure \fig{str1}: Star-triangle relation.}
\bigskip}

$$
\zeta_3=\zeta_1/\zeta_2.
\eqno(str)
$$

This relation is important in choosing appropriate couplings when deriving
functional equations.
For vertex operators, we shall use the star-triangle relation in an
iterated form shown in figure \reffig{str2}.
From figure \reffig{str1} we see that $H$ is of type $L$, $M$ of type $K$;
this is discussed further in section 2.4.
Notice that in each pair of operations the individual scalar factors $R$ are
mutually reciprocal, and therefore cancel.

\vbox{\bigskip
\centerline{\epsfbox{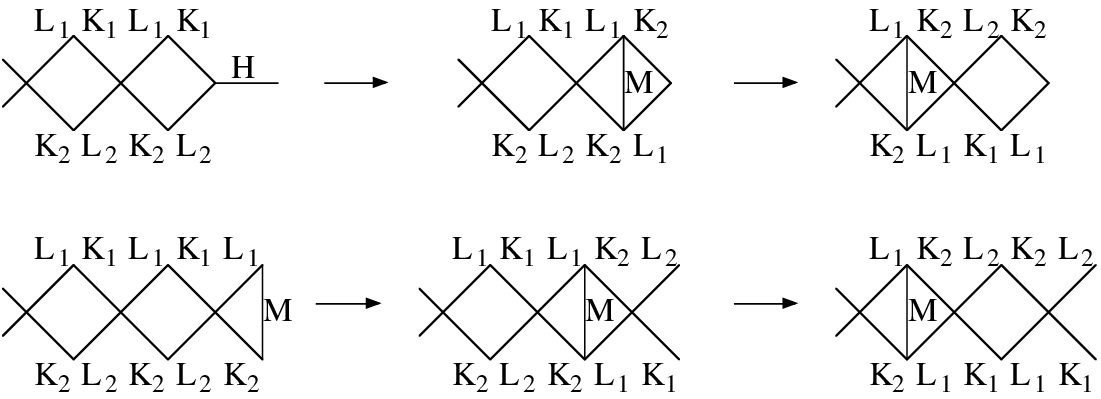}}
\bigskip
\centerline{Figure \fig{str2}: Iterated star-triangle relations.}
\bigskip}

\subsection{Corner transfer matrices}
A \CTM\ is defined to be the partition function
of a quadrant of the lattice, as shown in figure \reffig{ctm}.
The direction of transfer (action as an operator) is shown by an arrow. 
There are two sectors, \NS\ ($NS$) and \R\ ($R$) according as the
centre of the lattice is taken at a single spin site or not;
thus $\AN\zeta$ is diagonal with respect to the central spin $\epsilon$.
We fix the boundary spins to $+$.

\vbox{\bigskip
\centerline{\epsfbox{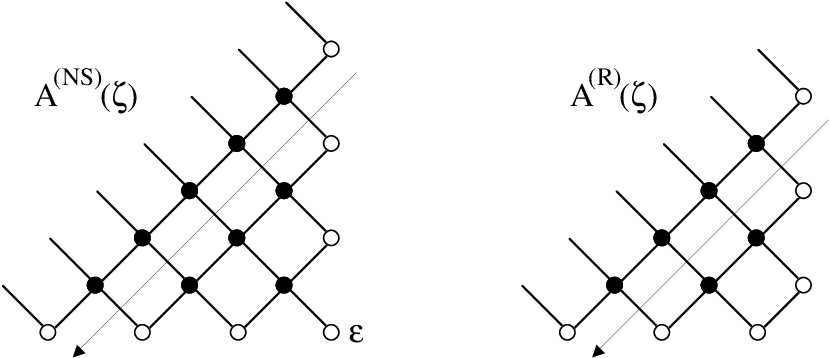}}
\bigskip
\centerline{Figure \fig{ctm}: Corner transfer matrices for the $NS$ and $R$
sectors.}
\bigskip}

The really remarkable properties of the \CTM s emerge in the thermodynamic
limit. 
In this limit, we expect that the entries of the \CTM s will scale as the
partition function per coupling raised to the power of the number of couplings,
and we assume that this factor is divided out to give a convenient finite limit.
 
These normalised infinite \CTM s are generated by spin chain operators $\HN$ and
$\HR$:
$$
\AN\zeta=\zeta^{\HN},
\qquad
\AR\zeta=\zeta^{\HR}.
$$
From this it is evident that the \CTM s are symmetric and that they satisfy the
group property
$$
\AN{\zeta_1}\AN{\zeta_2}=\AN{\zeta_1\zeta_2},
\qquad
\AR{\zeta_1}\AR{\zeta_2}=\AR{\zeta_1\zeta_2}.
\eqno(group)
$$
The eigenvectors of $\HN$ and $\HR$ span the \NS\ and \R\ sector, respectively.
Moreover, the \NS\ sector divides into two subsectors labelled by the value of
the central spin $\epsilon$.
Since it is straightforward to diagonalise $\HN$ and $\HR$ using fermions
\cite{Dav88,TP89}, analytic formulae may be given using properties of elliptic
functions.
In particular, the eigenvalue spectra are equally spaced, although we do not
make explicit use of these facts in this paper.

Finally we note that one may define \CTM s for the other three quadrants of the
lattice, North-East, South-East, South-West.
They are simply related to the one we have defined (North-West) by
$$
A_{\rm NE}(\zeta)=A'_{\rm NW}(x/\zeta),
\qquad
A_{\rm SE}(\zeta)=A_{\rm NW}(\zeta),
\qquad
A_{\rm SW}(\zeta)=A'_{\rm NW}(x/\zeta).
$$
provided that $\zeta$ retains its meaning as the ratio of horizontal and
vertical rapidities.
The prime denotes the transpose (which reverses the direction of transfer);
this is the form in which we construct $A_{\rm NE}$ from $A_{\rm NW}$ in
the non-symmetric case needed in section 6.

\subsection{Vertex operators}
The central technique of this paper is to employ ``vertex operators'' which map
between (intertwine) the \NS\ and \R\ sectors.
We are concerned with the physical interpretation of vertex operators
as semi-infinite row transfer matrices $V_\epsilon(\zeta)$ and
$W_\epsilon(\zeta)$, and in the functional relations which they satisfy.  
The two types of operators are shown in figure \reffig{vos}.
Since we shall often rotate these operators to other orientations, it is
important to note that they retain their underlying definition as
$V_\epsilon(K,L)$, $W_\epsilon(K,L)$, from which they derive their dependence
on $\zeta$ via \ref{KLdef}.
One sees immediately from the crossing symmetry that the vertex operators
satisfy
$$
V'_\epsilon(\zeta)=W_\epsilon(x/\zeta),
\qquad
W'_\epsilon(\zeta)=V_\epsilon(x/\zeta).
$$

\vbox{\bigskip
\centerline{\epsfbox{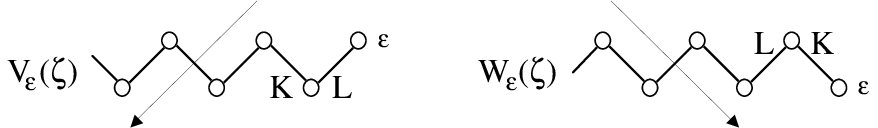}}
\bigskip
\centerline{Figure \fig{vos}: 
Vertex operators $V_\epsilon(\zeta)$ and $W_\epsilon(\zeta)$.}
\bigskip}

Consider now the left hand side of figure \reffig{vmat}, which shows the
composition of the action of $\AN\zeta$ with $V_\epsilon(\zeta)$.
For a finite system this simply gives a \R\ corner, 
but in the infinite case we must be more careful; $\AN\zeta$ maps the \NS\
sector into itself, after which $V_\epsilon(\zeta)$ maps it to the \R\ sector.
So, on the right-hand side of the equation, we have inserted a trivial operator
which makes the same mapping just by re-labelling the spins (the dashed lines 
indicate simple identification). 
However this operator is the anisotropic limit of $V_\epsilon(\zeta)$ at
$\zeta=1$, so we obtain the functional relation
$$
V_\epsilon(\zeta)\AN\zeta=\AR\zeta V_\epsilon(1).
$$
By eliminating $V_\epsilon(1)$ using the group property of \CTM s, we obtain the
general ``boost property''
$$
V_\epsilon(\zeta_1\zeta_2)\AN{\zeta_1}=\AR{\zeta_1}V_\epsilon(\zeta_2).
\eqno(boost:a)
$$
This boost property of \CTM s has been the subject of some investigation
\cite{SW83,Th86}, but its real utility only emerged recently
\cite{DFJMN93,JMN93,Dav94}.

\vbox{\bigskip
\centerline{\epsfbox{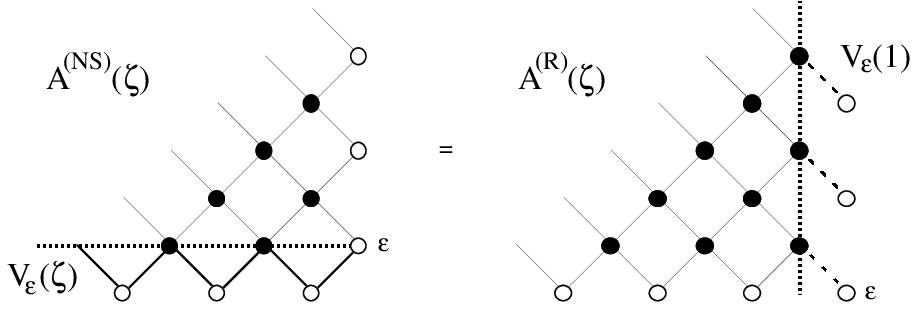}}
\bigskip
\centerline{Figure \fig{vmat}: Mapping from \NS\ to \R\ sectors using
$V_\epsilon(\zeta)$.}
\bigskip}

One can write a similar relation for the second type of vertex operator, which
we shall denote $W_\epsilon(\zeta)$.
The picture is shown in figure \reffig{wmat}, and an analogous argument gives
$$
W_\epsilon(\zeta_1\zeta_2)\AR{\zeta_1}=\AN{\zeta_1}W_\epsilon(\zeta_2).
\eqno(boost:b)
$$
Alternatively one can note that the two equations \ref{boost} are the transpose
of each other.
However, in the non-symmetric case of section 6, this will relate the action
of $A_{NE}$ and $A_{NW}$ on the vertex operators.

\vbox{\bigskip
\centerline{\epsfbox{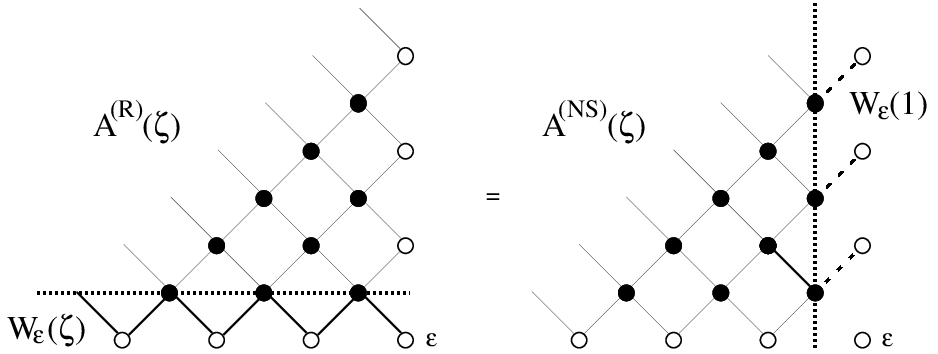}}
\bigskip
\centerline{Figure \fig{wmat}: Mapping from \R\ to \NS\ sector using
$W_\epsilon(\zeta)$.}
\bigskip}

Now consider the products $V_\epsilon(\zeta)W_\epsilon(\zeta')$ and
$W_\epsilon(\zeta)V_\epsilon(\zeta')$ which stand on the left hand side of
figure \reffig{commvw}. 
Using the star-triangle relation in the form shown in figure \reffig{str2},
we may move couplings $H$ or $M$ repeatedly from right to left, thereby
interchanging the Ising couplings, and with them the rapidity variables. 
Because of the fixed spin boundary conditions this will eventually cease to
have any further effect beyond multiplication by the simple factor 
${\rm e}^M$.
Thus we have ``commutation relations'', although it is the
rapidities rather than the operators which are interchanged.
This is required by the fact that they intertwine the \NS\ and \R\ sectors.

Consider first the product 
$F_\epsilon(\zeta,\zeta')=V_\epsilon(\zeta)W_\epsilon(\zeta')$, 
we have
$$
R{\rm e}^{M(\zeta/\zeta')}
F_\epsilon(\zeta,\zeta')
=\sum_{\epsilon'}{\rm e}^{H(\zeta/\zeta')\epsilon\epsilon'}
F_{\epsilon'}(\zeta',\zeta),
\eqno(vocomm:a)
$$
where $R$ comes from the first application of the star-triangle relation.

\vbox{\bigskip
\centerline{\epsfbox{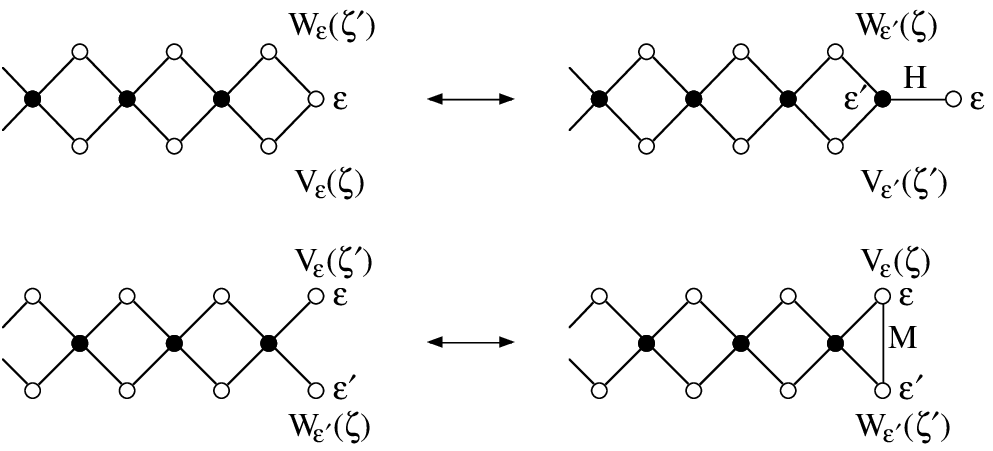}}
\bigskip
\centerline{Figure \fig{commvw}: Commutation of vertex operators.}
\bigskip}

\np
Typically we shall require combinations of the form
$F_{+}(\zeta,\zeta')\pm F_{-}(\zeta,\zeta')$, for which we have
$$
\eqalign{
R{\rm e}^{M(\zeta/\zeta')}(F_{+}(\zeta,\zeta')+F_{-}(\zeta,\zeta'))
=\cosh H(\zeta/\zeta')
(F_{+}(\zeta',\zeta)+F_{-}(\zeta',\zeta))\cr
R{\rm e}^{M(\zeta/\zeta')}(F_{+}(\zeta,\zeta')-F_{-}(\zeta,\zeta'))
=\sinh H(\zeta/\zeta')
(F_{+}(\zeta',\zeta)-F_{-}(\zeta',\zeta))\cr}
$$
When we are dealing with expectation values rather than operators, we shall
take quotients of such factors, as a result of which $R$ will
play no further part.
From the star-triangle relation one sees that 
$$
\tanh H(\zeta)=\tanh L(\zeta),
$$ 
the formula for which was given in \ref{tanhH}.

The case of the product 
$F_{\epsilon\epsilon'}(\zeta,\zeta')=W_\epsilon(\zeta)V_{\epsilon'}(\zeta')$
is similar, and we find
$$
{\rm e}^{M(\zeta/\zeta')}F_{\epsilon\epsilon'}(\zeta,\zeta')
={\rm e}^{M(\zeta/\zeta')\epsilon\epsilon'}
F_{\epsilon\epsilon'}(\zeta',\zeta).
\eqno(vocomm:b)
$$
For this formula $\tanh M(\zeta)=\tanh K(\zeta)$.
We shall, however, only use this relation in the case that $\epsilon=\epsilon'$,
for which we have the simple result that the rapidities may be interchanged
without introducing any factors.

%
%
%               Section 3
%
%
\section{Bulk magnetization}

As a preliminary to the calculation of surface and corner magnetizations
in later sections, we will develop the method, and the underlying assumptions,
by deriving the Onsager result for the bulk magnetization of the Ising model. 

\vbox{\bigskip
\centerline{\epsfbox{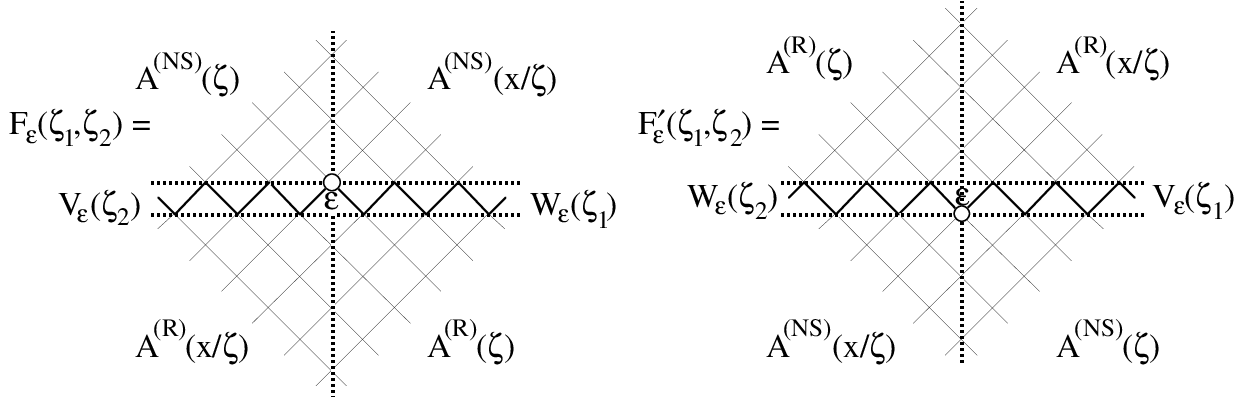}}
\bigskip
\centerline{Figure \fig{defect}: Lattices with a defect line.}
\bigskip}

\subsection{Basic properties}
Consider a lattice with a defect line as shown in figure \reffig{defect}.
The pictures represent the unnormalised expectation values
$$
\eqalign{
F_\epsilon(\zeta_1,\zeta_2)
&=\tr(\AR{x}V_\epsilon(\zeta_2)\AN{x}W_\epsilon(\zeta_1)),\cr
F'_\epsilon(\zeta_1,\zeta_2)
&=\tr(\AN{x}W_\epsilon(\zeta_2)\AR{x}V_\epsilon(\zeta_1)),\cr}
$$ 
where we have combined pairs of \CTM s, using the group property \ref{group}, to
eliminate the bulk rapidity.
The relation between $F$ and $F'$ is a simple reflection about the horizontal
axis. This interchanges the couplings $K$ and $L$, which is equivalent to
$\zeta\to x/\zeta$, so
$$
F_\epsilon(\zeta_1,\zeta_2)
=F'_\epsilon(x/\zeta_1,x/\zeta_2).
$$
This can be obtained directly from the definitions since $\tr(X')=\tr(X)$ for
any matrix $X$.

It is easy to see that the dependence on $\zeta_1$ and $\zeta_2$ is only
through the ratio $\zeta_1/\zeta_2$.
This is because these expectation values are matrix elements of a spin
operator between asymptotically degenerate maximal eigenvectors of the
horizontal row transfer matrices above and below the spin.    
But, because of the star-triangle relation, all eigenvectors of the row
transfer matrices except the line of dislocation are independent of rapidity
while the latter can only depend on the ratio $\zeta_1/\zeta_2$ \cite{BaxBk}. 
For the homogeneous case the magnetization is
$$
M_0
=\left({F_{+}(\zeta,\zeta)-F_{-}(\zeta,\zeta)\over
F_{+}(\zeta,\zeta)+F_{-}(\zeta,\zeta)}\right)
=\left({F'_{+}(\zeta,\zeta)-F'_{-}(\zeta,\zeta)\over
F'_{+}(\zeta,\zeta)+F'_{-}(\zeta,\zeta)}\right),
$$
Therefore we define the function $G(\zeta_1/\zeta_2)$ as
$$
G(\zeta_1/\zeta_2)
=\left({F_{+}(\zeta_1,\zeta_2)-F_{-}(\zeta_1,\zeta_2)\over
F_{+}(\zeta_1,\zeta_2)+F_{-}(\zeta_1,\zeta_2)}\right),
\eqno(Gdef)
$$
similarly for $G'(\zeta_1/\zeta_2)$, and 
$$
M_0=G(1)=G'(1).
\eqno(bulkmag)
$$

Finally, we shall require an inversion relation, namely
$$
G(\zeta)G(-\zeta)=1,
\eqno(inversion)
$$
from which the normalisation of $G(\zeta)$ may be determined.
It follows from
$$
F_\epsilon(-\zeta_1,\zeta_2)
=\rho\epsilon F_\epsilon(\zeta_1,\zeta_2),
$$
where $\rho=i^{n_K-n_L}$ and $n_K$, $n_L$ are the number of bonds of type
$K$ and $L$ in the vertex operator $W_\epsilon(\zeta_1)$, regarded now as finite
but large. 
To derive it, note that the transformation $\zeta_1\to-\zeta_1$ induces
$K\to K+i\pi/2$, $L\to L-i\pi/2$ in the operator
$W_\epsilon(\zeta_1)$ of figure \reffig{defect}, and the Boltzmann weights
transform as $\exp(\pm K)\to\pm(i)\exp(\pm K)$, 
$\exp(\pm L)\to\pm(-i)\exp(\pm L)$.
$F_\epsilon(\pm\zeta_1,\zeta_2)$ are sums over configurations of the system.  
For the ground state term the relation is obvious; after replacing
$\zeta_1$ by $-\zeta_1$, $\rho$ accounts for the phase factors $(\pm i)$
and the sign changes if $\epsilon=-1$ since one pair of spins
are not aligned in that case. 
But all other terms in the sum differ from the ground state only by having an
even number of further spin pairs of $W_\epsilon(\zeta_1)$ which are not
aligned; so every term in $F_\epsilon$ satisfies the same relation. 
From that, the result \ref{inversion} follows.

\vbox{\bigskip
\centerline{\epsfbox{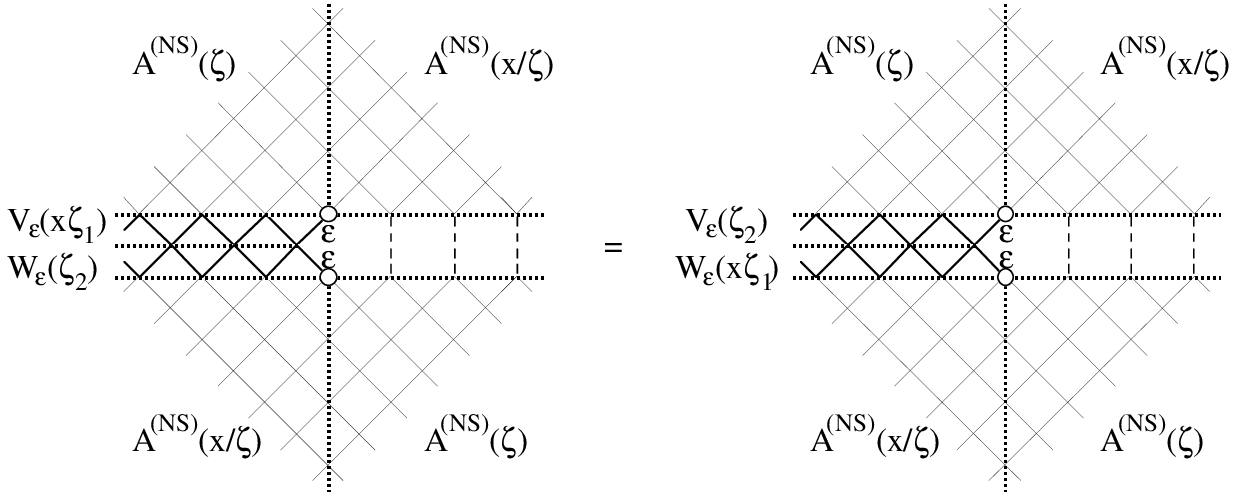}}
\bigskip
\centerline{Figure \fig{bulkde1}: Steps in deriving equation \ref{bulkde1}.}
\bigskip}

\subsection{Functional equation}
Deriving the functional ($q$-difference) equation involves quite a few steps
which we describe in detail.
Start with the function $F'_\epsilon(\zeta_1,\zeta_2)$, and perform the
following steps.

\point(i) 
Rotate the vertex operator $V_\epsilon(\zeta_1)$ 18\0\ anticlockwise, using the
boost property together with the cyclic property of the trace, to get the left
hand picture of figure \reffig{bulkde1}.
As a result, $\zeta_1\to x\zeta_1$.
\point(ii)
Interchange the rapidities using \ref{vocomm:b}, to give equality with the right
hand picture.
\point(iii)
Rotate $W_\epsilon(x\zeta_1)$ 18\0\ anticlockwise to get the
lattice which defines $F_\epsilon(x^2\zeta_1,\zeta_2)$.
\np
As a result we find
$$
F'_\epsilon(\zeta_1,\zeta_2)
=F_\epsilon(x^2\zeta_1,\zeta_2).
\eqno(bulkde1)
$$

For the second relation we proceed as in figure \reffig{bulkde2}:
\point(i) 
Starting with $F_\epsilon(\zeta_1,\zeta_2)$, rotate the vertex operator
$W_\epsilon(\zeta_1)$ 18\0\ anticlockwise to get the left hand picture of
figure \reffig{bulkde2}.
Again we have $\zeta_1\to x\zeta_1$.
\point(ii)
Interchange the rapidities, this time using \ref{vocomm:a}, to get to the
the right hand picture.
This requires that we use the coupling $H(\zeta_2/x\zeta_1)$.
\point(iii)
Rotate $V_\epsilon(x\zeta_1)$ 18\0\ anticlockwise to recover the
lattice which defines $F'_\epsilon(x^2\zeta_1,\zeta_2)$.
\np
Putting all of this together, we get the functional equations
$$
RF_\epsilon(\zeta_1,\zeta_2)
=\sum_{\epsilon'}{\rm e}^{H(\zeta_2/x\zeta_1)\epsilon\epsilon'}
F'_{\epsilon'}(x^2\zeta_1,\zeta_2).
\eqno(bulkde2)
$$
Finally, by substituting either of \ref{bulkde1,bulkde2} into the other, we
get uncoupled functional equations for either of the two functions.

\vbox{\bigskip
\centerline{\epsfbox{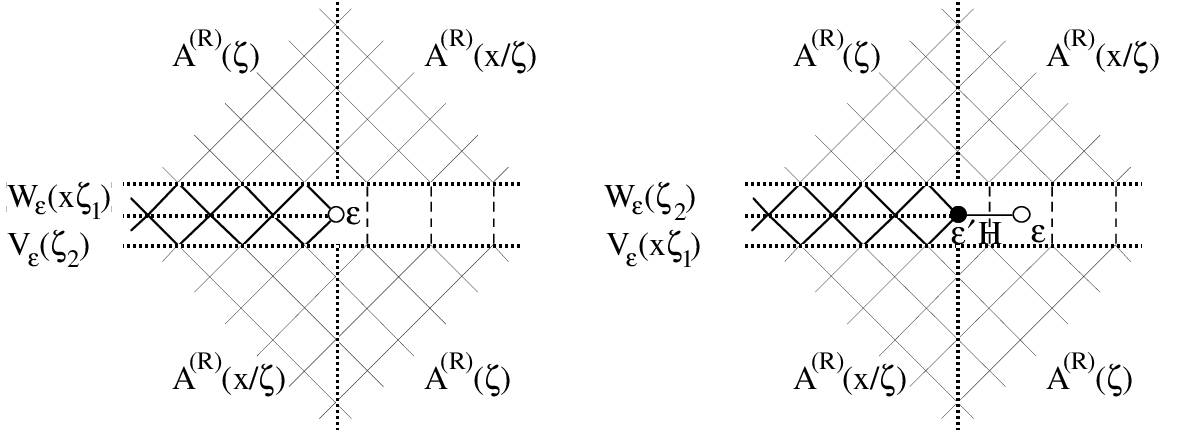}}
\bigskip
\centerline{Figure \fig{bulkde2}: Steps in deriving equation \ref{bulkde2}.}
\bigskip}

Changing to the functions $G(\xi)$, $G'(\xi)$ removes the dependence on
the factor $R$ which comes from the star-triangle relation, and gives
functional equations in a single variable.
We also set $x=q^{1/2}$ to get
$$
\eqalign{
G(\xi)
&=\tanh H({1/q^{1/2}\xi})G(q^2\xi),\cr
G'(\xi)
&=\tanh H({1/q^{3/2}\xi})G'(q^2\xi).\cr}
$$

\subsection{Solution of functional equation}
We shall only discuss the solution for $G(\xi)$.
Using the formula \ref{tanhH}, we obtain the basic functional equation,
$$
{G(\xi/q^2)\over G(\xi)}
={(\xi/q;q^2)(q^3/\xi;q^2)(-\xi;q^2)(-q^2/\xi;q^2)\over
(-\xi/q;q^2)(-q^3/\xi;q^2)(\xi;q^2)(q^2/\xi;q^2)}.
$$

For the purpose of solving functional equations, note that the functions
$$
f(\xi)=\ip(p^{\alpha+2}\xi;p^2,q^2),
\qquad
g(\xi)=\ip(p^\alpha/\xi;p^2,q^2),
\eqno(solnde:a)
$$
satisfy
$$
{f(\xi/p^2)\over f(\xi)}
=\ip(p^\alpha\xi;q^2),
\qquad
{g(\xi/p^2)\over g(\xi)}
={1\over\ip(p^\alpha/\xi;q^2)}.
\eqno(solnde:b)
$$
Here we employ the double infinite products
$$
\ip(z;p_1,p_2)
=\prod_{n_1,n_2=0}^\infty
(1-zp_1^{n_1}p_2^{n_2}).
$$

In the present case this gives
$$
G(\xi)=
{\ip(q\xi;q^2,q^2)\ip(-q^3/\xi;q^2,q^2)
 \ip(-q^2\xi;q^2,q^2)\ip(q^2/\xi;q^2,q^2)\over
 \ip(-q\xi;q^2,q^2)\ip(q^3/\xi;q^2,q^2)
 \ip(q^2\xi;q^2,q^2)\ip(-q^2/\xi;q^2,q^2)}.
$$
Notice that a possible normalising factor depending on $k$ is fixed (to unity)
by the inversion relation $G(\xi)G(-\xi)=1$ of equation \ref{inversion}. 
Substituting now into \ref{bulkmag}, we arrive at the formula
$$
M_0=G(1)=
{\ip(-q^3;q^2,q^2)\ip(q;q^2,q^2)\over
 \ip(q^3;q^2,q^2)\ip(-q;q^2,q^2)},
$$
where half the terms already cancelled.
Noting further that the remaining ratios fit into the patterns of
\ref{solnde}, we arrive at Onsager's original result
$$
M_0
={\ip(q;q^2)\over\ip(-q;q^2)}
=(1-k^2)^{1/8}.
\eqno(bulkm)
$$
So, although the form $(1-k^2)^{1/8}$ is simple, the infinite product carries
a message about the underlying structure of the problem.
It is interesting to note that Yang derived precisely this infinite product in
his original paper in which he proved the Onsager formula \cite{Yang52}.

\subsection{Uniqueness of solution}
The fundamental assumption of the method is that the function $G(\xi)$ is
meromorphic in $\xi$ and that its analytic
properties are determined by the analytic properties of the Ising couplings
as parametrized in \ref{KLdef}.
That is, it is a doubly periodic function of $u$. 
Even so, the functional equations by themselves determine their solution only to
within a ``pseudo-constant''; that is, a solution of $G(\xi)=G(q^2\xi)$. 
Such a function is doubly periodic in the variable $u$ so it is fully
determined by the structure of its poles and zeros.

Our basic hypothesis is that the analytic structure of $G(\xi)$, which
is actually the quotient of two functions $F_\pm$, is completely determined by
the factor $\tanh H$ which occurs in the equation for that quotient. 
In this case the solutions will only consist of the infinite product of those
factors, which also will automatically impose the normalisation up to a
constant depending only on $k$.
This constant was determined from the inversion relation \ref{inversion}.
The calculations, and the arguments employed, are exactly parallel to those of
\cite{JMN93} for the spontaneous polarization of the eight-vertex model.
This approach also has some precedent, although not so explicitly stated, in
Baxter's book \cite{BaxBk}.  
For example, the working of pp231-236, to obtain the free energy
of the eight-vertex model, is exactly the solution of a functional equation
(10.8.24) of the type we have just solved, and the solution is the infinite
product of just those terms which determine the relation.

In the case just cited \cite{BaxBk}, arguments are presented that the solution
actually has the appropriate meromorphic properties.
Actually, for the Ising model it is possible to construct an explicit analytic
expression for $G(\xi)$ as an integral \cite{FJMMN93}, using the fermionic
description of the \CTM s \cite{Dav88,TP89}, and check that
it does have the correct structure.

For the results which we present in the remainder of this paper we know of
no such integral expressions, but there are many points of contact with previous
results, some analytic, many numerical, and in all cases we have carefully
verified that there is exact agreement. 
It is also the case that all of our results can be viewed as special cases of a
single function. 
But the result for a $(10)$ surface has long been known \cite{MW67}, moreover
one of our results for a $(10)$ corner has recently been obtained independently
by solving integral equations \cite{AT94}.
Each of these checks is for the values of the meromorphic function
obtained here along a smooth curve in the complex plane, which is very strong
evidence for the validity of our approach.

%
%
%               Section 4
%
%
\section{Surface magnetizations}

In this section we generalize the method to a lattice with a $(11)$ surface.
This will allow us to calculate the magnetizations in the first and second
surface row. 

\subsection{Boundary reflection}
The essence of the bulk calculation is a rotation of vertex operators
through a total of 36\0.
For a system with a boundary this is not possible, but instead one can use a
procedure in which the vertex operators are rotated back and forth, being
``reflected'' from the boundary after a change of the rapidity.
This mechanism forms the basis of all further calculations. 

We are interested in simple free boundaries and we now show how this case can
be handled in an elementary way.
Consider the system shown in figure \reffig{edge1}.
Free boundaries require that we sum over the edge spins.
The effect of this is manifested through terms of the form
$$
\sum_\sigma{\rm e}^{(K\sigma_1+L\sigma_2)\sigma}
=2(\cosh K\cosh L+\sigma_1\sigma_2\sinh K\sinh L),
$$
where $\sigma$ is a boundary spin, and $\sigma_1$, $\sigma_2$ the two second
row spins joined directly to it by $K$ and $L$.
Now the crucial point is that the right hand side is completely symmetric with
respect to $K$, $L$ and also $\sigma_1$, $\sigma_2$.
The interchange of all pairs $K$, $L$ in a vertex operator which is adjacent
to a free edge is effected by $\zeta\to x/\zeta$; this gives
the following simple construction:

{\narrower\noindent\it 
a vertex operator adjacent to a free edge can have its rapidity
``reflected'' off the boundary via the transformations
$V_\epsilon(\zeta)\leftrightarrow V_\epsilon(x/\zeta)$,
$W_\epsilon(\zeta)\leftrightarrow W_\epsilon(x/\zeta)$.\par}

\noindent
This avoids the complication of having alternating rapidities and ``double row
transfer matrices'' which are normally associated with the use of boundary
reflections and $K$ matrices \cite{YB95}.

\vbox{\bigskip
\centerline{\epsfbox{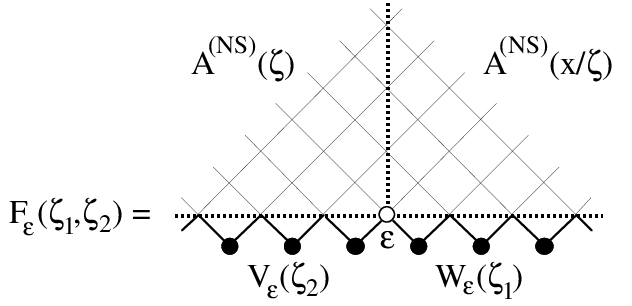}}
\bigskip
\centerline{Figure \fig{edge1}: System with a free boundary.}
\bigskip}

From this reflection property follow immediately two symmetries of the
expectation value shown in figure \reffig{edge1}:
$$
F_\epsilon(\zeta_1,\zeta_2)
=F_\epsilon(x/\zeta_1,\zeta_2)
=F_\epsilon(\zeta_1,x/\zeta_2).
\eqno(edgesymm)
$$
Finally, we note that the inversion relation \ref{inversion} still holds,
since the ground state configuration, even with free boundaries, still has all
spins except $\epsilon$ fixed to $+$, and the argument of section
3 then goes through. 

\subsection{Functional equations}
We now derive functional equations for the system shown in
figure \reffig{edge1}. 
In order that we can re-use these equations for a corner,
we temporarily replace $x$ by an arbitrary constant $p$ in the \CTM\
$\AN{x/\zeta}$, dealing instead with $\AN{p/\zeta}$.
The boundary reflection property, however, still involves the variable $x$.

We shall also consider directly the function $G(\zeta_1,\zeta_2)$,
as for the bulk magnetization.
For the first functional equation, we require the following steps (see
figure \reffig{edge2}):
\point(i)
Reflect the rapidity $\zeta_1$ off the right boundary; $\zeta_1\to x/\zeta_1$.
\point(ii)
Rotate the right vertex operator 18\0\ anti-clockwise so that the two are
adjacent; $x/\zeta_1\to px/\zeta_1$.
\point(iii)
Interchange the rapidities by multiplying $G$ with the first factor 
$\tanh H(\zeta_1\zeta_2/px)$.
\point(iv)
Reflect the rapidity $px/\zeta_1$ off the left boundary;
$px/\zeta_1\to \zeta_1/p$.
\point(v)
Interchange a second time multiplying with the second factor 
$\tanh H(\zeta_1/p\zeta_2)$.
\point(vi)
Rotate back 18\0\ to get the original system with $\zeta_1\to\zeta_1/p^2$.
\np
From these steps we obtain
$$
G(\zeta_1,\zeta_2)
=\tanh H(\zeta_1\zeta_2/px)
\tanh H(\zeta_1/p\zeta_2)
G(\zeta_1/p^2,\zeta_2).
\eqno(edgede:a)
$$

\vbox{\bigskip
\centerline{\epsfbox{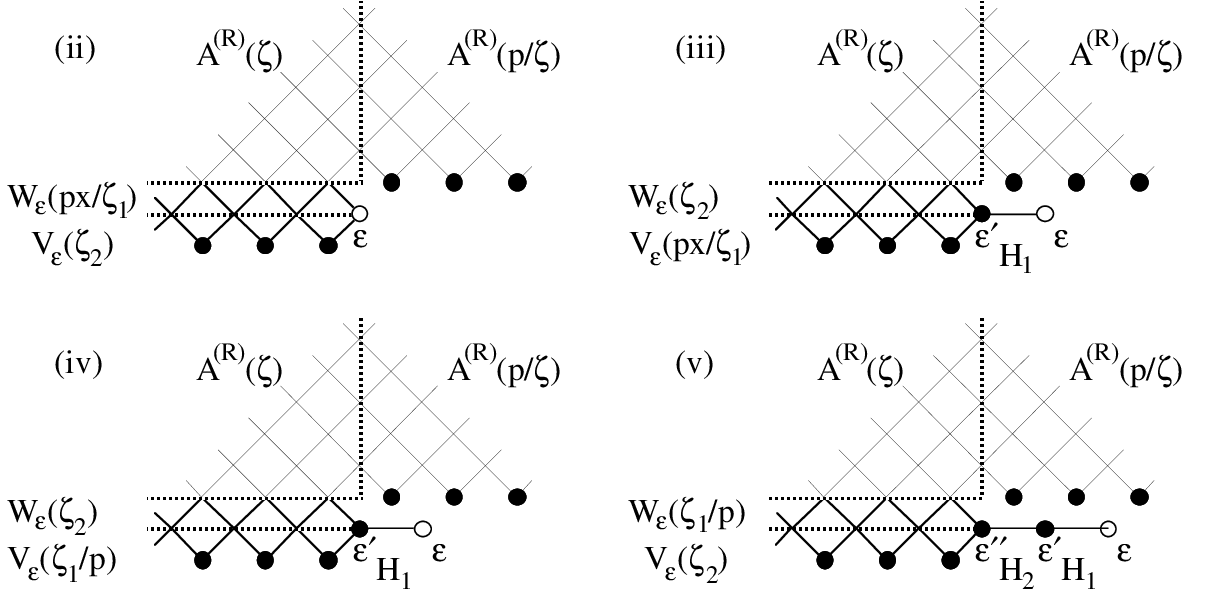}}
\bigskip
\centerline{Figure \fig{edge2}: Steps in deriving the functional
equation for a free surface.}
\bigskip}

\noindent
A similar sequence of moves involving the other vertex operator gives
$$
G(\zeta_1,\zeta_2)
=\tanh H(x/p\zeta_1\zeta_2)
\tanh H(\zeta_1/p\zeta_2)
G(\zeta_1,p^2\zeta_2).
\eqno(edgede:b)
$$

\subsection{Solution of the functional equations}
We have quite a number of relations which must be solved in a consistent way.
The first point to notice is that there are two factors which determine the
solution of the functional equations; one depends on the ratio
$\zeta_1/\zeta_2$ and the other on the product $\zeta_1\zeta_2$.
Therefore we use the Ansatz
$$
G(\zeta_1,\zeta_2)=\Phi(\zeta_1/\zeta_2)\Psi(\zeta_1\zeta_2).
\eqno(edgeansatz1)
$$
Imposing the symmetries \ref{edgesymm} on this Ansatz gives 
$$
\Phi(1/\xi)=\Phi(\xi),\qquad
\Psi(x/\xi)=\Psi(x\xi),\qquad
\Psi(\xi)=\Phi(\xi/x),
\eqno(edgeansatz2)
$$
so we need only find a single function $\Phi(\xi)$.
Equations \ref{edgede} give two functional equations for each of
$\Phi(\xi)$ and $\Psi(\xi)$, however it is easy to check that they are all
equivalent to the single equation
$$
\Phi(\xi)=\tanh H(\xi/p)\Phi(\xi/p^2).
\eqno(phide)
$$
Using \ref{solnde} and requiring that $\Phi(\xi)\Phi(-\xi)=1$, we obtain
$$
\Phi(\xi)
={\ip(pq^{1/2}/\xi;p^2,q^2)\ip(-pq^{3/2}\xi;p^2,q^2)
  \ip(-pq^{3/2}/\xi;p^2,q^2)\ip(pq^{1/2}\xi;p^2,q^2)\over
  \ip(-pq^{1/2}/\xi;p^2,q^2)\ip(pq^{3/2}\xi;p^2,q^2)
  \ip(pq^{3/2}/\xi;p^2,q^2)\ip(-pq^{1/2}\xi;p^2,q^2)}
\eqno(phi)
$$
Recall that, in order to use these results for an edge, we must set
$$
p=x=q^{1/2}.
$$

\subsection{First row magnetization}
The result for $G(\zeta_1,\zeta_2)$ gives directly the magnetization in the
second row, which will be discussed below.
But it also contains the result for the edge, because upon rotating
$V_\epsilon$ and $W_\epsilon$ into a vertical position, as shown in figure
\reffig{edge3}, the spin $\epsilon$ appears in the first row.
For a homogeneous lattice, the rapidities then have to equal $x/\zeta$.
Their original values before the rotation thus must be $\zeta_1=1$,
$\zeta_2=x$, as indicated also in the figure.
As a result, the edge magnetization is
$$
M_1=\Phi(1)\Phi(x).
$$
\vbox{\bigskip
\centerline{\epsfbox{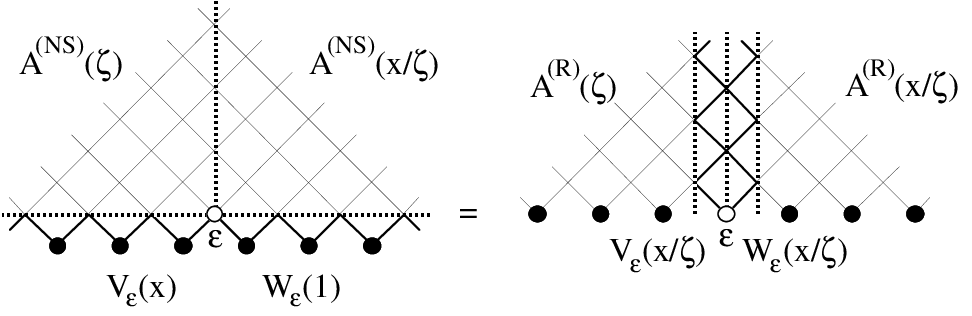}}
\bigskip
\centerline{Figure \fig{edge3}: Magnetization in first row.}
\bigskip}
\noindent
For both functions $\Phi$, the double infinite products can be
simplified into single infinite products as in section 3, giving
$$
M_1={\ip(q;q^2)^2\ip(q^{1/2};q)\over\ip(-q;q^2)^2\ip(-q^{1/2};q)}.
$$
According to \ref{bulkm} the first factor equals $(1-k^2)^{1/4}$.
The second one differs from it by the substitution $q\to q^{1/2}$, which
corresponds to a Landen transformation $k\to2k^{1/2}/(1+k)$ of the
modulus \cite{AS65}.
This leads to
$$
M_1=(1-k)^{1/2},
\eqno(edgem1)
$$
which corresponds exactly with the result in \cite{Pes85}, which was obtained
for the isotropic case.
We see, however, that $M_1$ is independent of the anisotropy, as it must be,
since it is the matrix element of a spin operator between the boundary state
and the maximal eigenvector of the row transfer matrix, both of which are
independent of $\zeta$.
This result also shows directly the surface exponent $\beta_s=1/2$. 

\subsection{Second row magnetization}
This is obtained by setting $\zeta_1=\zeta_2=\zeta$ in $G(\zeta_1,\zeta_2)$, so
that
$$
M_2=\Phi(1)\Phi(\zeta^2/x),
$$
which does depend on the anisotropy.
The dependence can, however, be simplified very much.
By reducing the double products to single products using \ref{solnde}, then
further to theta functions by \ref{theta}, one first shows that
$$
\Phi(\zeta^2/x)
=-i(k')^{1/2}{\sn(i(u+I'/2))\over\cn(i(u+I'/2))}.
$$
Using addition theorems for elliptic functions and substituting the Ising
couplings from \ref{KLdef} finally leads to
$$
M_2=(1-k)^{1/2}\coth(K+L).
\eqno(edgem2)
$$
This implies a simple relation between $M_2$ and $M_1$ which can also be
derived directly \cite{Pel96}.
The expression \ref{edgem2} is symmetric in $K$ and $L$, and $M_2$ exceeds $M_1$
for all finite positive $K$, $L$, as it should. 
Only in the anisotropic limit, where one diagonal line of couplings becomes
dominant, one has $M_2\to M_1$. 
This can already be seen from the initial expression for $M_2$ by setting
$\zeta=1$ and using $\Phi(x)=\Phi(1/x)$.

In the isotropic case ($\zeta^2=x$) one has
$$
M_2^{iso}=(1-k^2)^{1/2}=M_0^4,
\eqno(edgem3)
$$
which was already noted in \cite{Pes85}.
An analogous relation between bulk and surface quantities was found in
\cite{JKKMW95} for the eight-vertex model.
In that case the edge polarization (in the isotropic case) is the square of the
bulk polarization. 
One can show that this result implies \ref{edgem3} at the decoupling point.

The dependence of $M_2$ on the anisotropy is physically not obvious, but it is
exactly what one expects from an eigenvector calculation, for which
we would calculate the matrix element between the maximal
eigenvector of the row transfer matrix and the vector obtained by acting on the
boundary state with the row transfer matrix. 
However, such a calculation would be technically quite intricate
compared with the present method.

%
%
%               Section 5
%
%
\section{Corner magnetization}

In this section we consider 9\0\ corners formed by $(11)$ surfaces and
calculate the magnetization for three different spin positions. 

\vbox{\bigskip
\centerline{\epsfbox{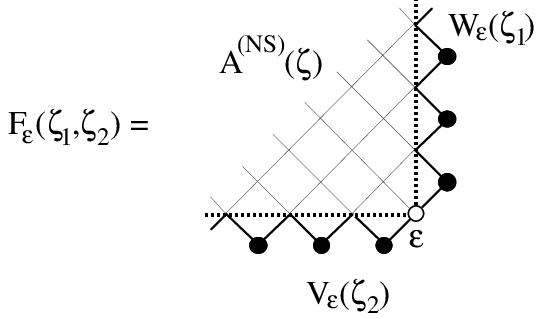}}
\bigskip
\centerline{Figure \fig{corner1}: Truncated corner with free
boundaries.}
\bigskip}

\subsection{Truncated corner}
We first consider the situation, shown in figure \reffig{corner1}, namely a
corner to which two vertex operators are attached, so that the tip
is missing.
This is the analogue of the edge geometry of figure \reffig{edge1}, and all
considerations of the previous section can be taken over immediately.
The only difference is that now the vertex operators are rotated only through
one \CTM, so that in deriving the formula for $G(\zeta_1,\zeta_2)$ one must set 
$$
p=\zeta.
$$
The boundary reflection property, however, still involves the variable $x$.
Therefore the equations \ref{edgede,edgeansatz1,edgeansatz2,phi} are unchanged
and the magnetization of the centre spin $\epsilon$ has the form
$$
M_2(\zeta)=\Phi(1)\Phi(q^{1/2}/\zeta^2).
\eqno(corner1)
$$
We postpone a discussion of the anisotropy dependence and first look at the
isotropic case, $\zeta=x^{1/2}=q^{1/4}$.
Then
$$
\eqalign{
M_2^{iso}
&=
{\ip(q^{3/4};q^{1/2},q^2)^4\ip(-q^{7/4};q^{1/2},q^2)^4\over
   \ip(-q^{3/4};q^{1/2},q^2)^4\ip(q^{7/4};q^{1/2},q^2)^4}\cr
&={\ip(q^{3/4};q^2)^4\ip(q^{5/4};q^2)^4\over
   \ip(q^{3/4};q^2)^4\ip(q^{5/4};q^2)^4}\cr}
$$
The products can be expressed as a ratio of Jacobian elliptic functions with
the help of \ref{theta},
$$
{\ip(q^{3/4};q^2)\ip(q^{5/4};q^2)\over
 \ip(-q^{3/4};q^2)\ip(-q^{5/4};q^2)}
=-(k')^{1/2}{\sn(3iI'/4)\over\cn(3iI'/4)}.
$$
Using Landen transformations to obtain the special values of $\sn$ and $\cn$
finally leads to the result
$$
M_2^{iso}=\left({1-\sqrt k\over1+\sqrt k}\right)
\left(\sqrt k+\sqrt{1+k}\right)^2.
\eqno(iso2)
$$

\vbox{\bigskip
\centerline{\epsfbox{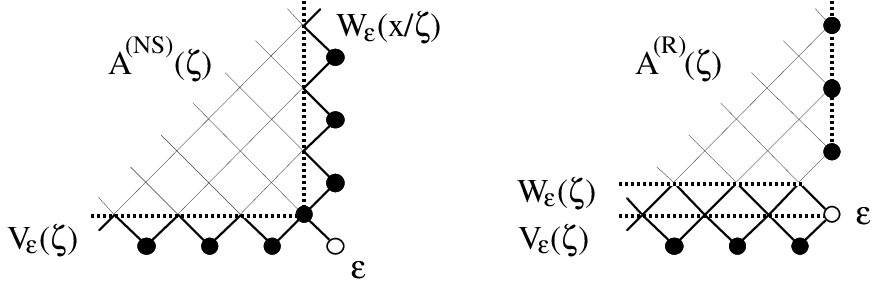}}
\bigskip
\centerline{Figure \fig{corner2}: \NS\ and \R\ corners.}
\bigskip}

\subsection{\NS\ corner}
This corner, shown on the left of figure \reffig{corner2}, is obtained by
adding a single spin with the coupling $K$ to the truncated corner.
The magnetization $M_2$ is not changed by this extra coupling while the
magnetization $M_1=\langle\epsilon\rangle$ is given by
$$
M_1(\zeta)
=\tanh K(\zeta)\cdot M_2(\zeta).
\eqno(cornerm1)
$$

Again, this simplifies very much for the isotropic case $\zeta=x^{1/2}=q^{1/4}$.
One can then either reduce the products introduced by $\tanh K(\zeta)$ as
before, or simply substitute $\tanh K(\zeta)=(\sqrt{k}+\sqrt{1+k})^{-1}$ to
obtain
$$
M_1^{iso}=\left({1-\sqrt k\over1+\sqrt k}\right)
\left(\sqrt k+\sqrt{1+k}\right).
\eqno(iso1)
$$

\subsection{\R\ corner}
This geometry is shown on the right of figure \reffig{corner2},
where it is also indicated that one can obtain it from  figure \reffig{corner1}
by rotating $W_\epsilon(\zeta_1)$ through the quadrant.
Therefore the magnetization $M_3$ of the spin $\epsilon$ follows by putting
$\zeta_1=1$, $\zeta_2=\zeta$ in $G(\zeta_1,\zeta_2)$ so that
$$
M_3(\zeta)=\Phi(\zeta)\Phi(\zeta/x),
$$
The isotropic case is readily extracted since the double infinite products
appear in a form which enables their reduction to single products by
\ref{solnde}, leading to
$$
M_3^{iso}=(1-k).
\eqno(iso3)
$$
This formula was already given in \cite{Pes85}.
By contrast, the other analytical results are new.
In particular those for $M_1^{iso}$ and $M_2^{iso}$ have been verified against
the numerical calculations carried out previously in \cite{Pes85,DP91}.

\subsection{Critical behaviour}
From the results \ref{iso2,iso1,iso3} one reads off directly the corner
exponent $\beta_c=1$ of the isotropic system.
However, for the corner considered here, $\beta_c$ is not a constant but
depends on the anisotropy (as it depends in general on the opening angle of the
corner).
This dependence can be calculated in a straightforward way from the exact
result \ref{corner1}.
While there is no general simplification of the double infinite products in
that case, it is not difficult to find the asymptotic form near the critical
point ($k\to1$) and hence evaluate $\beta_c$. 

The function $\Phi(\zeta)$ consists of factors which have the form
$$
\phi(\xi,\eta)=
{\ip(\xi;p^2,q^2)\ip(-\eta;p^2,q^2)\over
 \ip(-\xi;p^2,q^2)\ip(\eta;p^2,q^2)}.
$$
For $k\to1$ one has \cite{AS65} $I\to\infty$, $I'\approx\pi/2$, so that $q$ as
well as $\zeta$, and therefore the quantities $p$, $\xi$, $\eta$, all approach
one.
Therefore we set
$$
q={\rm e}^{-\epsilon},
\quad
p={\rm e}^{-\alpha\epsilon},
\quad
\xi={\rm e}^{-2x\epsilon},
\quad
\eta={\rm e}^{-2y\epsilon},
%\eqno(qprime)
$$
with $\epsilon\approx-\pi^2/\ln(1-k)\to0$ for $k\to1$, whereupon
$$
\ln\phi(\xi,\eta)
=\sum_{m,n=0}^\infty
\ln\left\{{\tanh(m+n\alpha+x)\epsilon\over\tanh(m+n\alpha+y)\epsilon}\right\}.
$$
For the asymptotic form we replace the double sum with a double integral and
apply the approximation
$$
{\tanh(s+\Delta)\over\tanh(s)}
\approx1+{2\Delta\over\sinh2s},
$$
Then
$$
\ln\phi(\xi,\eta)
\approx{2(y-x)\over\alpha\epsilon}
\int\hskip-8pt\int_0^\infty{ds\,dt\over\sinh2(s+t)}
={\pi^2\over8\epsilon}\left({x-y\over\alpha}\right).
$$
More generally we see that
$$
\sum_{i=1}^n
\ln\phi(\xi_i,\eta_i)
\approx{\pi^2\over8\alpha\epsilon}
\sum_{i=1}^n(x_i-y_i).
$$
Applying this to $\Phi(\zeta)$ we have
$$
\Phi(\zeta)\approx\exp\left(-{\pi^2\over8\alpha\epsilon}\right),
\eqno(phi1)
$$
which is dependent on $p$ via $\alpha$, but not explicitly dependent
on the argument $\zeta$. 
Setting finally $p=\zeta$ fixes $\alpha=u/\pi$.
With the value of $\epsilon$ given above it follows that the corner exponent is
$$
\beta_c
={1\over4\alpha}
={\pi\over4u},
\eqno(beta1)
$$

Since one is near $k=1$, the parameter $u$ can be expressed from \ref{KLdef} as
$$
u=\arctan({1\over\sinh2L_c}),
$$
where $L_c$ is the critical coupling across the corner.
This is precisely the result one obtains by rescaling the anisotropic system
onto an isotropic one with effective opening angle $\theta=2u$, together with
the conformal prediction $\beta_c=\pi/2\theta$ for the isotropic case
\cite{BPP84,IPT93}.
The present direct calculation is the first analytical proof of this argument.
Because $\Phi(\zeta)$ in \ref{phi1} does not depend explicitly on $\zeta$, the
same result for $\beta_c$ follows from $M_3$, as it should.

Finally, a treatment of corners with other opening angles would just lead to
other values of $p$.
For example, the straight surface corresponds to $p=x=q^{1/2}$, so that
$\alpha=1/2$ and one recovers the surface exponent $\beta_s=1/2$.
Thus \ref{beta1} is the general result and $u$ indeed plays the r\^ole of an
effective angle.

\subsection{Magnetization curves}
In order to give an impression how the corner magnetization varies with
temperature for different values of the anisotropy, we have calculated $M_1$
numerically from equation \ref{cornerm1}.
The results are shown in figure \reffig{corner3} for five values of the ratio
$L/K$ of the couplings (rather than $\zeta$).
One can see that for large $L/K$ the magnetization curve rises only slowly as
the temperature is decreased.
This mirrors the almost one-dimensional character of a system with strong
couplings across the corner.
Conversely, $M_1$ increases rapidly for small $L/K$.
The variation of the critical behaviour as described by $\beta_c$ is just a
particular aspect of this general pattern.

\vbox{\bigskip
\centerline{\epsfbox{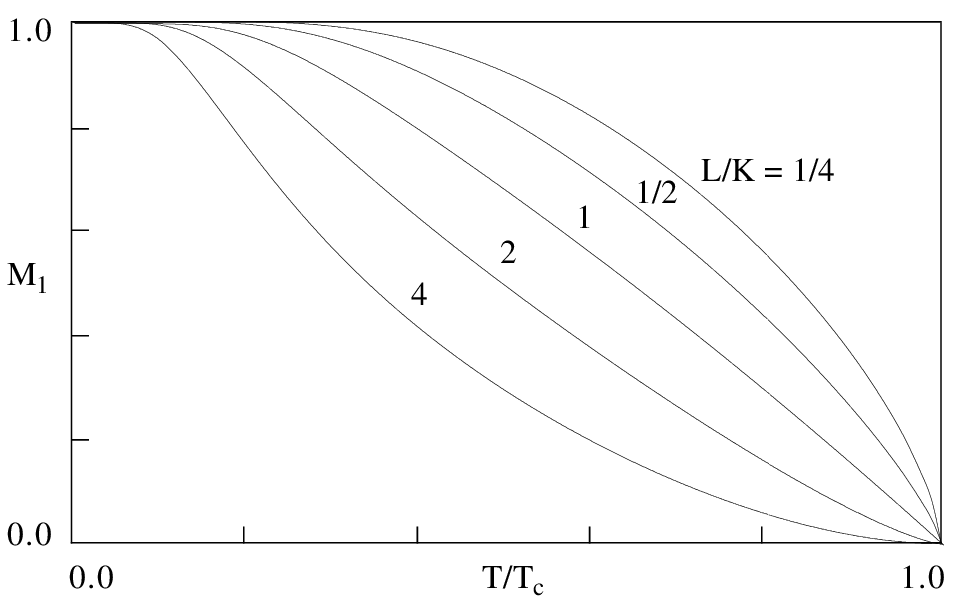}}
\bigskip
\centerline{Figure \fig{corner3}: Magnetization $M_1$ as a function of
temperature}
\centerline{for different values of anisotropy $L/K$.}
\bigskip}

%
%
%               Section 6
%
%f
\section{Surfaces in the principal direction}

The two previous sections have treated surfaces and corners formed by edges
taken in the diagonal $(11)$ direction.
In this section we shall show how to treat edges in the principal $(10)$,
$(01)$ directions using a layered system.

\vbox{\bigskip
\centerline{\epsfbox{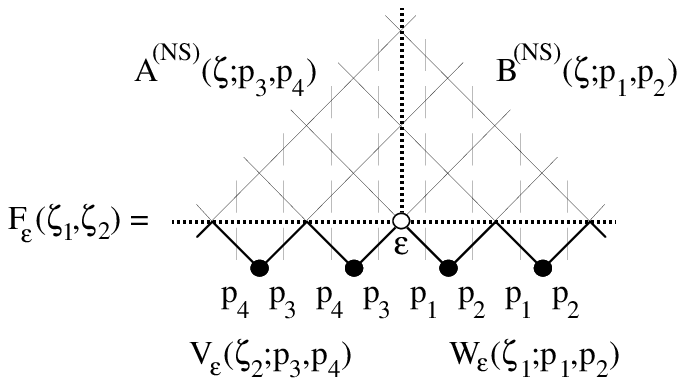}}
\bigskip
\centerline{Figure \fig{altedge1}: Layered system with a free boundary.}
\bigskip}

\subsection{Layered system}
Consider the system shown in figure \reffig{altedge1}.
It is essentially the same as figure \reffig{edge1}, except that we introduce
vertical rapidities $p_i$ which alternate between columns.
The meaning of the \CTM s and vertex operators which appear in the figure are
exactly as before, except that the couplings are calculated using the ratio of
horizontal to vertical rapidities, viz
$$
\eqalign{K&=K(\zeta/p),\cr L&=L(\zeta/p),\cr}
\qquad
\eqalign{\zeta&\in\{\zeta,\zeta_1,\zeta_2\},\cr p&\in\{p_1,p_2,p_3,p_4\}.\cr}
$$

We are particularly interested in some special choices of the $p$'s which
produce $(10)$ and $(01)$ surfaces.
These are shown in figures \reffig{altedge2} and \reffig{altedge3}.
Consider for example figure \reffig{altedge2:(a)}.
To get it from the general layered system we must choose $p_1$ and $p_4$ so
that the couplings which belong to their columns take the limits
$K(\zeta/p)\to\infty$, $L(\zeta/p)\to0$.
For this we must set $p_1=p_4=\zeta/x$.
This effectively removes half of the spins in the lattice, since each pair which
becomes identified by our choice is then equivalent to a single spin. 
Then we choose $p_2=p_3=1$ since we do not wish to alter the couplings in the
other two columns.
Finally we must set $\zeta_1=\zeta_2=\zeta$ to get the homogeneous lattice with
the $(10)$ surface shown in figure \reffig{altedge2:(a)}. 
These specializations will be made after the general solution has
been determined. 

\vbox{\bigskip
\centerline{\epsfbox{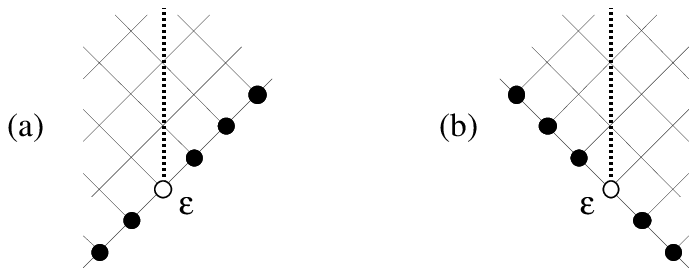}}
\bigskip
\centerline{Figure \fig{altedge2}: Surfaces with $(10)$ and $(01)$ free
 boundaries.}
\bigskip}

\np
Similar considerations give figure \reffig{altedge2:(b)}.  
So for figure
\reffig{altedge2} we need.
$$
\eqalignno{
&p_2=p_3=1,\qquad
p_1=p_4=\zeta/x,&(pvalues:a)\cr
&p_1=p_4=1,\qquad
p_2=p_3=\zeta,&(pvalues:b)\cr}
$$

\vbox{\bigskip
\centerline{\epsfbox{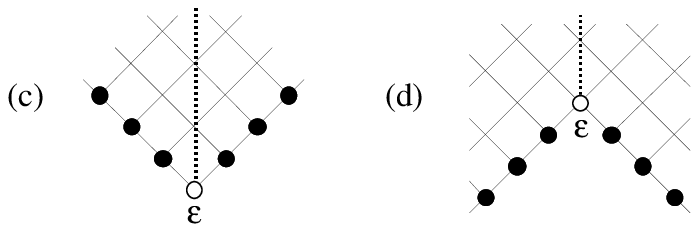}}
\bigskip
\centerline{Figure \fig{altedge3}: Corners with $(10)$ and $(01)$ free
 boundaries.}
\bigskip}

\noindent
Figure \reffig{altedge3} shows two types of corner, 9\0\ and 27\0.
For these the appropriate values are
$$
\eqalignno{
&p_2=p_4=1,\qquad
p_1=\zeta/x,\qquad
p_3=\zeta,&(pvalues:c)\cr
&p_1=p_3=1,\qquad
p_2=\zeta,\hphantom{/x}\qquad
p_4=\zeta/x,&(pvalues:d)\cr}
$$

We write $A(\zeta;p_3,p_4)$, $B(\zeta;p_1,p_2)$, $W_\epsilon(\zeta_1;p_1,p_2)$,
$V_\epsilon(\zeta_2;p_3,p_4)$ for the operators shown in figure
\reffig{altedge1}.
Just as in section 2, these operators are related by crossing symmetry.
Remembering that the adjoint reverses the direction of transfer, and that the
crossing symmetry must now be written as $\zeta/p\to xp/\zeta$, we see that
$$
\eqalign{
B(\zeta;p,p')&=A'(x/\zeta;1/p,1/p'),\cr
V_\epsilon(\zeta;p,p')&=W'_\epsilon(x/\zeta;1/p,1/p').\cr}
$$
We need the boost properties and commutation relations for these operators.
A short derivation is given in the appendix; here we state just the properties
used in this section.
Since we only have to rotate the operator $W_\epsilon(\zeta_1;p_1,p_2)$
through 18\0, we require the composition of two boosts. 
It is
$$
\eqalign{
&\AR{\zeta;p_3,p_4}\BR{\zeta;p_1,p_2}W_\epsilon(\zeta_1;p_1,p_2)\cr
&\qquad=W_\epsilon(x\zeta_1;p_3,p_4)\AN{\zeta;p_3,p_4}\BN{\zeta;p_1,p_2}.\cr}
\eqno(altboost)
$$

Next, we need the commutation relations \ref{vocomm} for the vertex
operators which have now aquired extra rapidities $p_i$.
The important point is that the 18\0\ rotation changes the vertical
rapidities $p_1$, $p_2$ to $p_3$, $p_4$, so they play no further r\^ole in
the commutation because they will match for the two operators which must be
multiplied; this means that equations \ref{vocomm} actually remain valid
with the same choice of couplings, which depend only on the horizontal
rapidities.

The last ingredient is the ``boundary reflection''.
Recall that this is simply the interchange of adjacent pairs $K(\zeta/p)$,
$L(\zeta/p')$ which are joined to a common spin on the free boundary, since
the sum is symmetric in both the variables and the other pair of spins.
This means that we want to make the replacements
$\zeta/p\to xp'/\zeta$, $\zeta/p'\to xp/\zeta$.
At first sight this brings the difficulty that the vertical rapidities have
changed place.
But we observe that the replacement may also be written as
$\zeta/p\to(xpp'/\zeta)/p$, $\zeta/p'\to(xpp'/\zeta)/p'$, which is equivalent
to a change in only the horizontal rapidity by $\zeta\to(xpp'/\zeta)$.
Thus we have the simple rule:

{\narrower\noindent\it 
a vertex operator adjacent to a free edge can have its rapidity
``reflected'' off the boundary via the transformations
$V_\epsilon(\zeta;p_1,p_2)\leftrightarrow 
V_\epsilon(xp_1p_2/\zeta;p_1,p_2)$,
$W_\epsilon(\zeta;p_3,p_4)\leftrightarrow 
W_\epsilon(xp_3p_4/\zeta;p_3,p_4)$.\par}

\subsection{Functional equations}
We now derive functional equations for the system shown in
figure \reffig{altedge1}. 
We shall re-use the figures and steps which were used in section 4, except
that we retain the original meaning of $x=q^{1/2}$ in the \CTM s.
For the first functional equation, we require the following steps (see
figure \reffig{edge2}):
\point(i)
Reflect the rapidity $\zeta_1$ off the right boundary:
$\zeta_1\to xp_1p_2/\zeta_1$.
\point(ii)
Rotate the right vertex operator 18\0\ anti-clockwise so that the two are
adjacent: $xp_1p_2/\zeta_1\to x^2p_1p_2/\zeta_1$.
\point(iii)
Interchange the rapidities by multiplying $G$ with the factor 
$\tanh H(\zeta_1\zeta_2/x^2p_1p_2)$.
\point(iv)
Reflect the rapidity $x^2p_1p_2/\zeta_1$ off the left boundary:
$x^2p_1p_2/\zeta_1\to p_3p_4\zeta_1/xp_1p_2$.
\point(v)
Interchange a second time multiplying with the factor 
$\tanh H(p_3p_4\zeta_1/xp_1p_2\zeta_2)$.
\point(vi)
Rotate back 18\0\ to get the original system with 
$\zeta_1\to\zeta_1/r^2$, where 
$$
r^2=x^2{p_1p_2\over p_3p_4}.
$$
\np
From these steps we obtain
$$
G(\zeta_1,\zeta_2)
=\tanh H(\zeta_1\zeta_2/x^2p_1p_2)
\tanh H(x\zeta_1/r^2\zeta_2)
G(\zeta_1/r^2,\zeta_2).
\eqno(altedgede:a)
$$
A similar sequence of moves involving the other vertex operator gives
$$
G(\zeta_1,\zeta_2)
=\tanh H(p_3p_4/\zeta_1\zeta_2)
\tanh H(x\zeta_1/r^2\zeta_2)
G(\zeta_1,r^2\zeta_2).
\eqno(altedgede:b)
$$
These equations, and equations \ref{altedgeansatz2} below,
generalize those of section 4, to which they reduce if all $p_i=1$.

\subsection{Solution of the equations}
We follow the procedure of section 4 by using the Ansatz
$$
G(\zeta_1,\zeta_2)=\Phi(\zeta_1/\zeta_2)\Psi(\zeta_1\zeta_2).
\eqno(altedgeansatz1)
$$
The boundary reflection property imposes symmetries which are a
generalisation of \ref{edgesymm}, namely
$G(\zeta_1,\zeta_2)=G(xp_1p_2/\zeta_1,\zeta_2)=G(\zeta_1,xp_3p_4/\zeta_2)$,
which manifest themselves as
$$
\Phi(1/p_3p_4\xi)=\Phi(p_1p_2\xi),\quad
\Psi(xp_1p_2/\xi)=\Psi(xp_3p_4\xi),\quad
\Psi(\xi)=\Phi(xp_1p_2/\xi).
\eqno(altedgeansatz2)
$$
As before we need only find a single function $\Phi(\xi)$, since equations
\ref{altedgede} are all equivalent to
$$
\Phi(\xi)=\tanh H(x\xi/r^2)\Phi(\xi/r^2).
$$
The solution may be written down by reference to equations \ref{phide,phi},
whereupon one sees that it is only required to make the substitutions
$p\to r$, $\xi\to x\xi/r=q^{1/2}\xi/r$ in order to obtain
$$
\Phi(\xi)
={\ip(r^2/\xi;r^2,q^2)\ip(-q^2\xi;r^2,q^2)
  \ip(-qr^2/\xi;r^2,q^2)\ip(q\xi;r^2,q^2)\over
  \ip(-r^2/\xi;r^2,q^2)\ip(q^2\xi;r^2,q^2)
  \ip(qr^2/\xi;r^2,q^2)\ip(-q\xi;r^2,q^2)}.
\eqno(altphi)
$$
The actual solution for any particular surface or corner is obtained by
substituting this result, together with the special values of the parameters
$p_i$, into \ref{altedgeansatz1,altedgeansatz2}.

\subsection{9\0\ corner}
The relevant parameters for this corner are given in \ref{pvalues:c} from which
$r^2=x=q^{1/2}$.
For the magnetisation, we have 
$$
M_c(\zeta)=\Phi(1)\Phi(1/\zeta),
\qquad
r=q^{1/4}.
\eqno(mag90)
$$
In the isotropic case, $\zeta=q^{1/4}$, we have already evaluated both
products; $\Phi(1)$ in section 4.4 and $\Phi(q^{-1/4})$ in section 5.1.
Reusing those results, we have
$$
\Phi(1)=\sqrt{1-k},
\qquad
\Phi(q^{-1/4})=\left({1-\sqrt k\over1+\sqrt k}\right)^{1/2}
\left(\sqrt k+\sqrt{1+k}\right),
$$
giving
$$
M^{iso}_c=(1-\sqrt{k})(\sqrt k+\sqrt{1+k}).
\eqno(mag90iso)
$$

We can show that the general formula \ref{mag90} agrees with the known result
of Kaiser and Peschel \cite{KP89,AT94},
$$
M_c=1-(\coth K-1)(\coth L-1)/2.
\eqno(KP)
$$
Using the simple identity  $(\coth K-1)^2=2(\cosh2K-\sinh2K)/(\cosh2K-1)$,
it may be written in terms of half-angles as
$$
M_c(\zeta)=(1-k)
{\dn(iu/2)^2-k\cn(iu/2)^2\over\dn(iu/2)(\dn(iu/2)+ik\sn(iu/2)\cn(iu/2))}.
\eqno(KP1:a)
$$
To see that this is the same as \ref{mag90}, first note that when we set
$r^2=q^{1/2}$ in $\Phi(1/\zeta)$, the double products simplify to elliptic
functions as follows:
$$
\eqalign{
\Phi(1/\zeta)&=
 {\ip(q^{1/2}\zeta;q^{1/2},q^2)\ip(-q^2/\zeta;q^{1/2},q^2)
  \ip(-q^{3/2}\zeta;q^{1/2},q^2)\ip(q/\zeta;q^{1/2},q^2)\over
  \ip(-q^{1/2}\zeta;q^{1/2},q^2)\ip(q^2/\zeta;q^{1/2},q^2)
  \ip(q^{3/2}\zeta;q^{1/2},q^2)\ip(-q/\zeta;q^{1/2},q^2)}\cr
&=
 {\ip(q^{1/2}\zeta;q^2)\ip(q\zeta;q^2)
  \ip(q/\zeta;q^2)\ip(q^{3/2}/\zeta;q^2)\over
  \ip(-q^{1/2}\zeta;q^2)\ip(-q\zeta;q^2)
  \ip(-q/\zeta;q^2)\ip(-q^{3/2}/\zeta;q^2)}\cr
&=
 -k'{\sn(i(u/2+I'/2))\sn(i(u/2+I'))\over
  \cn(i(u/2+I'/2))\cn(i(u/2+I'))}.\cr}
$$
Using addition theorems to eliminate the quarter period $I'$ gives the form
$$
M_c(\zeta)
=(1-k){\cn(iu/2)\dn(iu/2)-i(1+k)\sn(iu/2)\over
\dn(iu/2)(\cn(iu/2)-i\sn(iu/2)\dn(iu/2))}
\eqno(KP1:b)
$$
After a fair amount of further calculation, one can show that the two formulae
\ref{KP1:{a,b}} agree.

\subsection{Surface}
Using \ref{pvalues:a} we have now $r^2=x^2=q$, and for the
magnetisation,
$$ 
M_s(\zeta)=\Phi(1)\Phi(1/\zeta),
\qquad
r=q^{1/2}.
\eqno(mag180:a)
$$
Substituting $r^2=q$ in the formula for $\Phi$ and reducing the double products
to single products gives
$$
\Phi(1/\zeta)
={\ip(q\zeta;q^2)\ip(q/\zeta;q^2)\over
 \ip(-q\zeta;q^2)\ip(-q/\zeta;q^2)}
%=-ik'^{1/2}{\sn(i(u+2I')/2)\over\cn(i(u+2I')/2)}
=k'^{1/2}{\nd(iu/2)}.
\eqno(mag180:b)
$$

It is easy to reconcile equations \ref{mag180:{a,b}} with the formula of McCoy
and Wu \cite{MW67} for the magnetization at a $(10)$ surface, namely
$$
M_s
=\left({\cosh2K-\cosh2\hat{L}\over\cosh2K-1}\right)^{1/2},
\eqno(MW)
$$
where $\hat{L}$ is the dual coupling; $\cosh2\hat{L}=\coth2L$.
One simply substitutes from \ref{KLdef}, and uses half
angle formulae for the Jacobian elliptic functions, to obtain
$$
M_s
=(1-k^2)^{1/2}{\nd(iu/2)}.
$$

Finally, we note that choosing $\zeta_1=\zeta_2=\zeta'$ different from the
bulk rapidity $\zeta$ leads to the system shown in figure \reffig{AY}.
By summing out the spins in the first row one then obtains a $(10)$ surface
with modified couplings in the boundary, a problem treated in detail by Au-Yang
\cite{AY73}.

\vbox{\bigskip
\centerline{\epsfbox{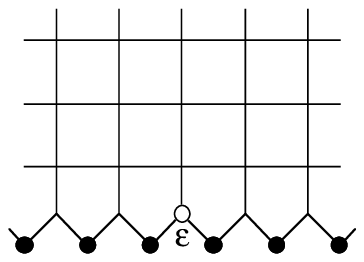}}
\bigskip
\centerline{Figure \fig{AY}: System with a modified $(10)$ boundary.}
\bigskip}

\subsection{27\0\ corner}
Using \ref{pvalues:d} we have now $r^2=x^3=q^{3/2}$, and for the
magnetisation,
$$ 
M_c=\Phi(1)\Phi(q^{1/2}/\zeta),
\qquad
r=q^{3/4}.
$$
However, there does not appear to be any simple reduction to Jacobian elliptic
functions in this case.
This is evident from the fact that the two bases in the doubly
infinite products, $q^{3/4}$ and $q^2$, are not related by Landen
transformation in the simple way they were for the 9\0\ and 18\0\ cases.
However, the asymptotic analysis of the section 5.4 is easily applied since
the function $\Phi(\xi)$ of this section is essentially the same as there,
and one sees that the critical exponent is
$$
\beta_c=1/3,
$$
independent of the anisotropy, as it must be.

%
%
%               Section 7
%
%
\section{Conclusion}

We have presented an approach which permits the calculation of order
parameters in Ising models with various kinds of simple boundaries.
The method uses \CTM s which already on geometrical grounds
are the natural tool for the shapes studied.
But while a previous attempt to use them was only partly successful
\cite{DP91}, the vertex operator technique leads to a very elegant and compact
solution.
Moreover it has a marked geometrical character which makes it easy to use.
We have deliberately presented it in some detail, because so far the method is
not widely known.
By contrast, we have not addressed questions of the infinite lattice limit
which is always involved implicitly when \CTM s are introduced.

The results obtained from the functional equations all have the form of
infinite products which reflect the spectrum of the \CTM s as well as the
matrix elements with the boundary state vectors.
In general the formulae are quite complicated.
But they are not difficult to derive, and we have seen that amazing
simplifications may be achieved which can lead to very simple final
expressions for the order parameters.
It can in fact be quite difficult to find these reductions, and in a way, when
found, the simple forms obscure the origin and structure of the results.

One can think of many applications to the Ising model beyond those studied
in this paper. 
Here we have not studied two-point correlation functions, or indeed more
general correlation functions, although they are
solutions of similar (but more complex) functional equations.
Other surface configurations could also be studied: for example, by
adding more vertex operators at the boundaries one can in principle treat
spins in deeper layers. 
Again, the layered system in section 6 has only been used for simple special
cases related to a particular problem, and certainly not investigated in any
generality.
In this context one should remember that \CTM s were
themselves developed by considering inhomogeneous systems, as were vertex
operators and their functional equations.
Now, in this paper, inhomogeneties have been used to rotate the surface
from the $(11)$ to the $(10)$ and $(01)$ directions.

The method of $q$-difference equations is not confined to the Ising model,
having been used in papers on the eight vertex model \cite{JMN93,Las94}, the
$Z_{n+1}\times Z_{n+1}$ generalized Heisenberg model \cite{Koy93}, and for the
\XYZ\ chain with a boundary \cite{JKKMW95}.
However the method is presented here in the language of spins for the first
time, and this simplifies matters considerably compared with using \ABF\ models,
or taking the eight-vertex model at the decoupling point.
As such it applies naturally to the chiral Potts model, although there are
formidable difficulties in that case due to the fact that the rapidities lie on
a high-genus Riemann surface.

\section*{Acknowledgements}

B. D. would like to thank the Deutscher Akademischer Austauschdienst
(DAAD) and the Freie Universit\"at Berlin for support and hospitality.

\numberby{}
\prefixby{A}
%
%
%               Appendix
%
%
\section*{Appendix}

\subsection*{Group property}
We need to extend the group property \ref{group} to include alternating
vertical rapidities.
The derivation is the same for both \NS\ and \R\ sectors, and for the $NW$ and
$NE$ corners.
Here we choose $\AN{\zeta;p,p'}$ and show that
$$
\AN{\zeta_1;p,p'}\AN{\zeta_2}=\AN{\zeta_1\zeta_2;p,p'},
\eqno(agroup)
$$
where $\AN{\zeta}=\AN{\zeta;1,1}$.
A simple derivation follows from the observation that $\AN{\zeta;p,p'}$ may
be regarded as the infinite product of {\it vertical} semi-infinite row
transfer matrices \cite{BaxBk},
$$
\AN{\zeta_1;p,p'}=\cdots
W_\epsilon(xp'/\zeta_1)V_\epsilon(xp/\zeta_1)
W_\epsilon(xp'/\zeta_1)V_\epsilon(xp/\zeta_1).
$$
Multiplying on the right by $\AN{\zeta_2}$ and invoking \ref{boost}
repeatedly (to commute the $V$ and $W$ through $A$) we arrive at the desired
result. 
One can write it also with a scalar normalisation factor, but we assume
that such factors are absorbed into the definition of the \CTM s, as for the
original group property \ref{group}.

\subsection*{Boost property}
We sketch the derivation of the boost property \ref{altboost}.
As in section 2, it is trivial to see that
$$
\eqalign{
V_\epsilon(\zeta;p_1,p_2)\AN{\zeta;p_1,p_2}
&=\AR{\zeta;p_1,p_2}V_\epsilon(1),\cr
W_\epsilon(\zeta;p_1,p_2)\AR{\zeta;p_1,p_2}
&=\AN{\zeta;p_1,p_2}W_\epsilon(1).\cr}
\eqno(aboost1)
$$
We have also the original boost property \ref{boost} for $V_\epsilon(\zeta)$ and
$W_\epsilon(\zeta)$, 
%$$
%\eqalign{
%V_\epsilon(\zeta_1\zeta_2)\AN{\zeta_1}
%&=\AR{\zeta_1}V_\epsilon(\zeta_2),\cr
%W_\epsilon(\zeta_1\zeta_2)\AR{\zeta_1}
%&=\AN{\zeta_1}W_\epsilon(\zeta_2).\cr}
%$$
which may be used, together with \ref{agroup}, to ``lift'' \ref{aboost1} to
$$
\eqalign{
V_\epsilon(\zeta_1\zeta_2;p_1,p_2)\AN{\zeta_1;p_1,p_2}
&=\AR{\zeta_1;p_1,p_2}V_\epsilon(\zeta_2),\cr
W_\epsilon(\zeta_1\zeta_2;p_1,p_2)\AR{\zeta_1;p_1,p_2}
&=\AN{\zeta_1;p_1,p_2}W_\epsilon(\zeta_2).\cr}
\eqno(aboost1)
$$
Alternatively, on can view this as an expression of $Z$-invariance \cite{Bax78}
applied to the {\it rotation} of a rapidity line through 9\0; in this process
the alternating vertical rapidities present in the horizontal row are absorbed
into the \CTM.

Taking the transpose of \ref{aboost1} gives the analogous property for
$\BN{\zeta;p,p'}$:
$$
\eqalign{
\BN{\zeta_1;p_1,p_2}W_\epsilon(\zeta_1\zeta_2;p_1,p_2)
&=W_\epsilon(x\zeta_2)\BR{\zeta_1;p_1,p_2},\cr
\BR{\zeta_1;p_1,p_2}V_\epsilon(\zeta_1\zeta_2;p_1,p_2)
&=V_\epsilon(x\zeta_2)\BN{\zeta_1;p_1,p_2}.\cr}
\eqno(aboost2)
$$
Notice the factor $x$ which comes from the crossing symmetry by which we
obtain the $\BN{\zeta;p,p'}$ corner from the $\AN{\zeta;p,p'}$.
The required result \ref{altboost} now follows from the composition of
\ref{aboost1} with \ref{aboost2}.

\section*{References}

\bibitem{ABF84} \jnlitem
{G. E. Andrews, R. J. Baxter and P. J. Forrester:
Eight-vertex SOS model and generalized Rogers-Ramanujan type identities}
{\JSP}{35}{(1984) 193--266}

\bibitem{AS65} \bkitem
{M. Abramowitz and I. A. Stegun}
{Handbook of mathematical functions}
{Dover, 8th ed. (1972)}

\bibitem{Abr71} \jnlitem
{D. B. Abraham: On the transfer matrix for the two-dimensional
Ising model}{Studies in Applied Mathematics}{50}{(1971) 71--88}

\bibitem{AT94} \jnlitem
{D. B. Abraham and F. Latr\'emoli\`ere:
Corner spontaneous magnetization}
{\PR}{E50}{(1994) R9--R11}
 \jnlitem {--- : Corner spontaneous magnetization}
{\JSP}{81}{(1995) 539--559}

\bibitem{AY73} \jnlitem
{H. Au-Yang: Thermodynamics of an anisotropic boundary of a two dimensional
Ising model}
{\JMP}{14}{(1973) 937--946}

\bibitem{Bax78} \jnlitem
{R. J. Baxter:
Solvable eight-vertex model on an arbitrary planar lattice}
{\PTRSL}{289}{(1978), 315-346}

\bibitem{BaxBk} \bkitem
{R. J. Baxter}{Exactly Solved  Models in Statistical Mechanics}
{Academic Press (1982)}

\bibitem{Bax85} \bkitem
{R. J. Baxter:
Exactly Solved  Models in Statistical Mechanics}
{Integrable systems in Statistical Mechanics VII}
{eds. G. M. d.Ariano, A. Monterosi and M. G. Rasetti, 
World Scientific, Singapore (1985) 5--63}

\bibitem{BPP84} \jnlitem
{M. N. Barber, I. Peschel and P. A. Pearce:
Magnetization at corners in two-dimensional Ising models}
{\JSP}{37}{(1984) 497--527}

\bibitem{Car83} \jnlitem
{J. L. Cardy: Critical behaviour at an edge}
{\JPA}{16}{(1983) 3617--3628}

\bibitem{Dav88} \jnlitem
{B. Davies: Corner transfer matrices for the Ising model}
{\PA}{154}{(1989) 1--20}

\bibitem{DP91} \jnlitem
{B. Davies and I. Peschel: 
Corner transfer matrices and corner magnetization for the Ising model}
{\JPA}{24}{(1991) 1293--1306}

\bibitem{Dav93}
{B. Davies: Corner Transfer Matrices and Quantum Affine Algebras}
{\JPA}{27}{(1994) 361--378} 

\bibitem{Dav94} \bkitem
{B. Davies: Infinite Dimensional Symmetry of corner transfer matrices}
{Confront\-ing the Infinite}
{ed. A. L. Carey et. al., World Scientific, Singapore (1995) 175--192}

\bibitem{DFJMN93} \jnlitem
{B. Davies, O. Foda, M. Jimbo, T. Miwa and A. Nakayashiki:
Diagonalization of the XXZ Hamiltonian by vertex operators}
{\CMP}{151}{(1993) 89--153}

\bibitem {DJKMO87} \jnlitem {E. Date, M. Jimbo, A. Kuniba, T. Miwa and M. Okado:
Exactly solvable SOS models: Local height probabilities and theta function
identities} {\NPB}{290}{(1987) 231--273}

\bibitem{FJMMN93} \jnlitem
{O. Foda, M. Jimbo, K. Miki, T. Miwa and A. Nakayashiki:
Vertex operators in solvable lattice models}{\JMP}{35}{(1994) 13--46}

\bibitem{IPT93} \jnlitem
{F. Igl\'oi, I. Peschel and L. Turban:
Inhomogeneous systems with unusual critical behaviour}
{\AP}{42}{(1993) 683--740}

\bibitem{JMN93} \jnlitem
{M. Jimbo, T. Miwa and A.  Nakayashiki:
Difference equations for the correlation functions
of the eight-vertex model}
{\JPA}{26}{(1993) 2199--2209}

\bibitem{JKKMW95} \jnlitem
{M. Jimbo, R. Kedem, H. Konno, T. Miwa and R. Weston:
Difference equations in spin chains with a boundary}
{\NPB}{448}{(1995) 429--456}

\bibitem{Koy93} \jnlitem
{Y. Koyama:
Staggered polarization of vertex models with $U_p(sl(n))$ symmetry}
{\CMP}{164}{(1994), 277-291}

\bibitem{KP89} \jnlitem
{C. Kaiser and I. Peschel:
Surface and corner magnetizations in the two-dimen\-sional Ising model}
{\JSP}{54}{(1989) 567--579}

\bibitem{Las94} \jnlitem
{M. Yu. Lashkevich:
Equations for correlation functions of eight-vertex model:
Ferromagnetic and disordered phases}
{\MPL}{B10}{(1996) 101--116}

\bibitem{MW67} \jnlitem
{B. M. McCoy and T. T. Wu:
Theory of Toeplitz determinants and the spin correlations of
the two-dimensional Ising model: IV}
{\PR}{162}{(1967) 436--475}

\bibitem{MW73} \bkitem
{B. M. McCoy and T. T. Wu}
{The two-dimensional Ising model}{Harvard University
Press; Cambridge Mass. (1967)}

\bibitem{Ons49} \jnlitem
{G. S. Rushbrooke: On the theory of regular solutions}
{\NC\ Supplement}{6}{(1949) 251--263 --- see discussion p. 261}

\bibitem{Pes84} \jnlitem
{I. Peschel: 
Surface magnetization in inhomogeneous two-dimensional Ising lattices}
{\PR}{B30}{(1984) 6783--6784}

\bibitem{Pes85} \jnlitem
{I. Peschel: 
Some more results for the Ising square lattice with a corner}
{\PLA}{110}{(1985) 313--315}

\bibitem{Pel96} \jnlitem
{A. Pelizzola: 
Exact boundary magnetization of the layered Ising model on
triangular and honeycomb lattices}
{\MPL}{B10}{(1996) 145--151}

\bibitem{TP89} \jnlitem
{T. T. Truong and I. Peschel: 
Diagonalisation of finite-size corner transfer matrices and related
spin chains}
{\ZPB}{75}{(1989) 119--125}

\bibitem{SW83} \jnlitem
{K. Sogo and M. Wadati:
Boost operator and its application to quantum Gelfand-Levitan
equation for Heisenberg-Ising chain with spin one-half}
{\PTP}{69}{(1983) 431--450}

\bibitem{Th86} \jnlitem
{H. B. Thacker:
Corner transfer matrices and Lorentz invariance on a lattice}
{\PD}{18}{(1986) 348--359}

\bibitem{Yang52} \jnlitem
{C. N. Yang: The spontaneous magnetization of a two-dimensional Ising model}
{\PR}{85}{(1952) 808--816}

\bibitem{YB95} \jnlitem
{C. M. Yung and M. T. Batchelor: Integrable vertex and loop models on the
square lattice with open boundaries via reflection matrices}
{\NPB}{435}{(1995) 430--462}

\listreferences

\bye